\documentclass[%
 aip,
 amsmath,amssymb,
preprint,%
]{revtex4-2}

\usepackage{graphicx}% Include figure files
\usepackage{dcolumn}% Align table columns on decimal point
\usepackage{bm}% bold math
\usepackage{color}
\usepackage[normalem]{ulem}

\usepackage[utf8]{inputenc}
\usepackage[T1]{fontenc}
\usepackage{mathptmx}

\newcommand\scalemath[2]{\scalebox{#1}{\mbox{\ensuremath{\displaystyle #2}}}}

\begin{document}

\preprint{AIP/123-QED}

%\title[A theory for the flow of chemically-responsive polymer solutions]{A theory for the flow of chemically-responsive polymer solutions: equilibrium and shear-induced phase separation}
\title{A theory for the flow of chemically-responsive polymer solutions: equilibrium and shear-induced phase separation}
% Force line breaks with \\

\author{Marco De Corato}
 %\altaffiliation[Current address ]{Physics Department, XYZ University.}%Lines break automatically or can be forced with \\
%\author{B. Author}%
 \email{mdecorato@unizar.es}
\affiliation{ 
Aragon Institute of Engineering Research (I3A), University of Zaragoza, Zaragoza, Spain%\\This line break forced with \textbackslash\textbackslash
}%

\author{Marino Arroyo}
 %\homepage{http://www.Second.institution.edu/~Charlie.Author.}
\affiliation{%
Universitat Polit\`{e}cnica de Catalunya-BarcelonaTech, 08034 Barcelona, Spain%\\This line break forced% with \\
}%
\affiliation{%
Institute for Bioengineering of Catalonia (IBEC), The Barcelona Institute of Science and Technology (BIST), Baldiri Reixac 10-12, 08028 Barcelona, Spain; %\\ Centre Internacional de Metodes Numerics en Enginyeria (CIMNE), 08034 Barcelona, Spain%\\This line break forced% with \\
}%
\affiliation{%
Centre Internacional de M\`{e}todes Num\`{e}rics en Enginyeria (CIMNE), 08034 Barcelona, Spain%\\This line break forced% with \\
}%

\date{\today}% It is always \today, today,
             %  but any date may be explicitly specified

\begin{abstract}
Chemically-responsive polymers are macromolecules that respond to local variations of the chemical composition of the solution by changing their conformation, with notable examples including polyelectrolytes, proteins and DNA. The polymer conformation changes can occur in response to changes to the pH, the ionic strength or to the concentration of a generic solute that interacts with the polymer. These chemical stimuli can lead to drastic variations of the polymer flexibility and even trigger a transition from a coil to a globule polymer conformation. In many situations, the spatial distribution of the chemical stimuli can be highly inhomogeneous, which can lead to large spatial variations of polymer conformation and of the rheological properties of the mixture. In this paper, we develop a theory for the flow of a mixture of a solute and chemically-responsive polymers. The approach is valid for generic flows and inhomogeneous distributions of polymers and solutes. To model the polymer conformation changes introduced by the interactions with the solute, we consider the polymers as linear elastic dumbbells whose spring stiffness depends on the solute concentration. We use the Onsager's variational formalism to derive the equations governing the evolution of the variables, which unveils novel couplings between the distribution of dumbbells and that of the solute. Finally, we use a linear stability analysis to show that the governing equations predict an equilibrium phase separation and a distinct shear-induced phase separation whereby a homogeneous distribution of solute and dumbbells spontaneously demix. Similar phase transitions have been observed in previous experiments using stimuli-responsive polymers and may play an important role in living systems.
%\marino{[Very nice but the first part of the abstract reads more like an introduction. Can be shortened.]}
\end{abstract}

\maketitle

\section{Introduction}
%START BY SAYING THAT POLYMER CONFORMATION CHANGES A LOT AS A FUNCTION OF THE SOLVENT. SAY THAT THIS IS WHAT WE CALL CHEMICALLY-RESPONSIVE POLYMERS.

%REVISE THE FIGURES FONTS

Chemically-responsive polymers constitute a class of macromolecules that can change their conformation in response to chemical stimuli such as changing the chemical composition of a suspension \cite{GIL_2004}. These types of polymers is commonly encountered in several biomedical and industrial applications \cite{Jeong_2002,Stuart_2010}. For instance,  polyelectrolytes  can change their conformation in response to changes of the pH or of the ionic strength of the solution. Change of local salt concentration screens the charges along the backbone of the polyelectrolytes thus making the polymer chain more flexible and promoting a transition from an extended semiflexible chain configuration to a flexible and coiled one  \cite{De_Gennes_1976,Odijk_1977,Fixman_1982,DOBRYNIN_2005}. The gradual change in the polymer conformation has been observed in many experiments that measured the persistence length of the chain at different salt concentrations using optical and magnetic tweezers \cite{Marko_1995,Baumann_1997,Guilbaud_2019}. A more extreme type of chemically-responsive behavior is displayed by linear poly(N-isopropylacrylamide)(PNIPAM) polymers. Experiments and simulations measuring the extension of PNIPAM chains at different salt and solute concentrations \cite{Schild_1991,Zhang_2005,Mukherji_2014} found a transition between a coiled conformation and a globular one. Such coil to globule transition is often accompanied by a phase separation of the polymeric phase. 

Not surprisingly, the conformation of  many bio-polymers that are crucial for the correct functioning of living organisms is sensitive to small changes to the chemical composition of its environment. This feature can be exploited by eukaryotic cells and bacteria to actively control the function of these important macromolecules \cite{Erdel_2018,Parker_2021}. For instance, DNA and RNA can change their conformation depending on the local concentration of binding proteins \cite{Ha_1999,Barbieri_2012}. Similarly, proteins can change their conformation in response to chemical stimuli such as changes of pH or ionic strength \cite{Chou_2020}. In some cases, the conformational changes of these macromolecules are associated to phase separation and organization of these biopolymers into membrane-less compartments, which carry out critical tasks for cell survival \cite{Hyman_2014,Brangwynne_2015,Shin_2017,Choi_2020}. 
Finally, a last class of chemically-responsive polymers are those that weakly bind in the presence of molecules that bridge different groups along the polymer backbone, thus changing its conformation \cite{Diamant_2000,Levy_2019}.  

In general, the conformational changes of chemically-responsive polymers affect the rheological properties of the suspension  \cite{rubinstein2003polymer}. For instance, in the case of polyelectrolytes, increasing the salt concentration leads to more compact configurations that are not easily stretched by straining flows. Indeed, experiments in shear flow show that the viscosity decreases as the salt concentration is increased, which is a signature of the smaller dissipation generated by the macromolecules in their compact configuration  \cite{Rubinstein_1994,Dobrynin_1995, Chen_2020, Chen_2021}. Besides changing the viscosity of the solution, conformational changes can also impact its viscoelastic properties because more compact macromolecules relax faster to their equilibrium shape than those that are in an extended conformation  \cite{rubinstein2003polymer}. This qualitative picture is confirmed by small-amplitude oscillatory shear and microfluidics experiments showing a change of the relaxation time as a function of the concentration of a solute  \cite{Del_Giudice_2017,Turkoz_2018,Zhang_2020}. 

%In the recent years there has been a great effort in incorporating chemical reactions into rheology [McKinley JNNFM, Germann JoR, Greeks JoR, Ianniruberto JoR, Larson BD JoR, Mike Cates Macromolecules]

%The flow of chemically-responsive polymers has been often studied in rheometric flows where the shear rate is and the concentration of polymers and solutes are homogeneous. Interestingly, in these type of flows both shear-banding and shear-induced demixing have been observed. While shear banding instabilities are linked to a nonmonotonic relation between the shear stress and the shear rate \cite{Fielding_2014,Divoux_2016}, shear-induced phase demixing is driven by the coupling between gradient of stresses and gradients of polymer density \cite{Helfand_1989,Larson_1992,Cromer_2013,Cromer_2014}. [THETWO SENTENCES ABOVE HAS TO BE CHANGED]  Since the polymers respond to changes of local chemical composition it is interesting to investigate if coupling between solute and polymer conformation could promote the instability. Another relevant question arises in the simulation of inhomogeneous flows of chemically-responsive polymers where the solute concentration changes in space leading to inhomogeneous rheological properties. How should one modify the viscoelstic constitutive equations to include the solute-dependent rheological properties? A naive answer would be to simply use spatially-variable viscosity and relaxation time. However, this procedure might ignore important couplings between the gradients of solute and gradients of polymer density and conformation.

Standard theories for polymer solutions such as the Flory-Huggins theory  \cite{rubinstein2003polymer} capture the interaction between a polymer and a solvent whose concentration is homogeneous and has been used to predict the phase separation, the coil-globule transition and the dependence of the elastic modulus of polymer solutions depending on the solvent quality. However, this theory cannot describe polymers in a solvent interacting with solute molecules since in general solute concentration can be heterogeneous as a consequence of chemical reactions, external gradients, solute or polymer advection or other nonequilibrium processes that are ubiquitous in living systems. Even in the simplest case of a shear flow with a homogeneous polymer concentration, flow instabilities can lead to  gradients of concentration transverse to the flow direction \cite{Yanase_1991,Larson_1992,Fielding_2014,Cromer_2013,Cromer_2014,Divoux_2016,Germann_2019,Burroughs_2020}.

Another relevant question arises in the simulation of inhomogeneous flows of chemically-responsive polymers where the solute concentration changes in space leading to inhomogeneous rheological properties. How should one modify the viscoelastic constitutive equations \cite{larson2013constitutive} to include the solute-dependent rheological properties? A naive answer would be to simply use spatially-variable viscosity and relaxation time. However, this procedure might ignore important couplings between the gradients of solute and polymer density. Furthermore, since the polymer conformation depends on the local chemical composition, it is interesting to investigate if unexpected couplings between solute and polymer conformation could occur. 

In this paper, we address these points by proposing a thermodynamically self-consistent model for the isothermal flow of a dilute mixture of solvent, solute and chemically-responsive polymers. Our model is applicable to general flows with spatially-variable strain rates, not necessarily unidirectional, and inhomogeneous solute and polymer concentrations. To develop a general model of chemically-responsive polymers, we assume that conformation changes driven by the presence of a solute can be modeled as a change of the polymer stiffness. By using the Onsager's variational formalism, we derive  systematically the set of equations that govern the flow of a mixture of solute and chemically-responsive polymers. These equations reveal couplings between the gradients of solute, the gradients of polymer density and the gradients of polymer conformation, which are not present in previous viscoelastic models. Finally, we show that the solute-polymer coupling leads to distinct mechanisms of demixing of a homogeneous solution, equilibrium and shear-induced.

\section{Model for polymer-solute interactions}

\begin{figure}[h!]
\centering
\includegraphics[width=1.0\textwidth]{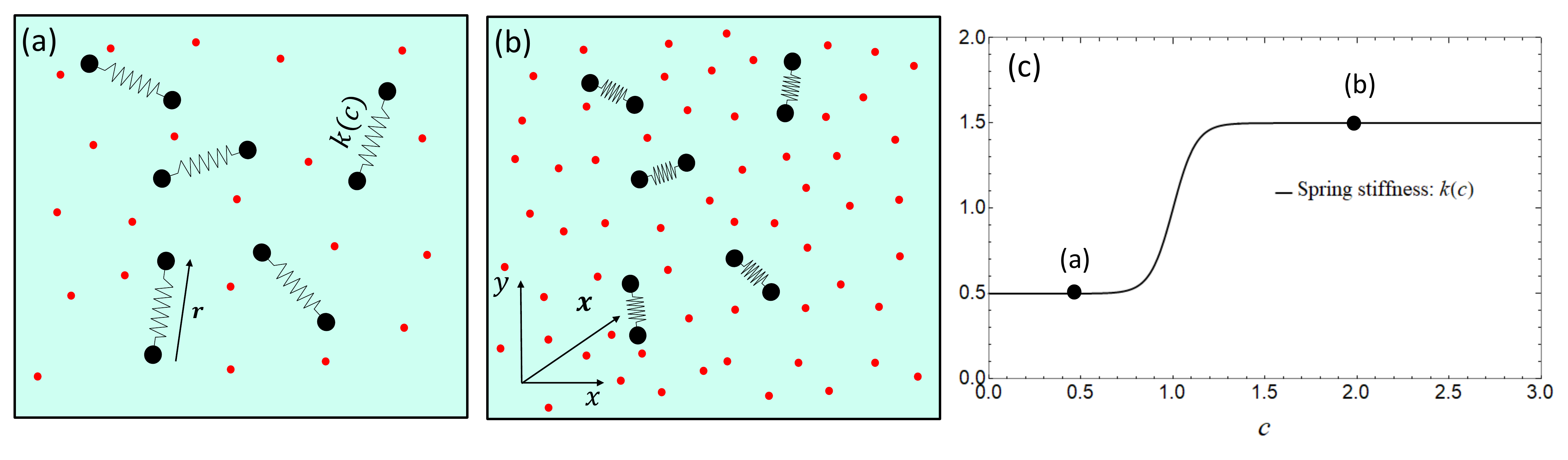}
\caption{Schematics of a mixture of chemically-responsive polymers and solute in a solvent. The polymers are modeled as dumbbells, which are characterized by their position vector $\boldsymbol{x}$, end-to-end vector $\boldsymbol{r}$ and stiffness $k(c)$, with $c$ the solute number density. The solute molecules, typically much smaller than the polymers, are represented as red circles. Panel (c) illustrates qualitatively the case of dumbbells whose spring stiffness increases with the local solute concentration. In panel (a) the concentration of solute is small, the spring is soft and the dumbbells are in an extended configuration. By increasing the solute concentration as shown in panel (b), the spring becomes stiffer and the dumbbells are in a contracted configuration. In this way, we model conformation changes of polymers introduced by the spatial variations of a solute. }
\label{fig1}
\end{figure}

To describe chemically-responsive polymer solutions in general flow conditions and heterogeneous concentrations, we make the simple choice of modeling  polymers as dumbbells. i.e.~two beads connected by a spring, see Figure \ref{fig1}. The vector joining the two spheres of the dumbbell represents the end-to-end vector of the real polymer. The dumbbell model is a good approximation of polymer suspensions in dilute and semidilute regimes where the polymer chains are not entangled with each other \cite{Larson_2005}. We assume that the stiffness of the spring connecting the two beads of a dumbbell depends on the concentration of solute. By doing so, we provide a simple, yet general, model for the conformation change of polymers introduced by the presence of a solute. Since the extension of the dumbbells  represents the end-to-end vector of the polymer, by considering a stiffness that depends on the solute concentration we can model chemically-induced conformational changes. For instance, the coil to globule transition of polymers caused by the interactions with a solute can be modeled through a steep change of the spring stiffness with the concentration of solute, see Figure \ref{fig1}. %The reduction of persistence length, gyration radius and viscosity observed in experiments at increasing solute concentration can all be modeled through an increase of the dumbbell stiffness with an increase concentration of solute. \marino{[the two sentences above largely overlap in meaning]}

We assume that the dumbbells and the solute are suspended in a solvent of viscosity $\eta$ and that the solution is incompressible and isothermal. We are interested in spatially-extended systems in which the concentration of polymers and of solute can change in space. Therefore, we characterize the distribution of dumbbells inside a volume, $\Omega$, through the function $\psi(\boldsymbol{x},\boldsymbol{r},t)$ , where $\boldsymbol{x}$ denotes the position of the center of a dumbbell in space, $\boldsymbol{r}$ denotes its end-to-end vector and $t$ is the time, see Figures \ref{fig1}(a-b). The local number density of dumbbells at a given point in space is given by the integral over the end-to-end distance space:
\begin{equation}\label{dumbb_conc}
n_\text{d}(n_\text{d}, \boldsymbol{x},t)= \int_{\boldsymbol{r}} \psi(\boldsymbol{x},\boldsymbol{r},t) \, d\boldsymbol{r} \, \, .
\end{equation}
The distribution of solute is simply characterized by its number density $c(\boldsymbol{x},t)$. We assume that the elastic energy stored by each dumbbell molecule is given by a linear spring:
\begin{equation}\label{elast_energ}
U = \frac{1}{2} \, k(c) \, \vert\boldsymbol{r}\vert^2\, \, .
\end{equation}
The elastic energy, given by Eq. \eqref{elast_energ}, has an entropic origin that reflects the resistance of the polymer chains to be stretched away from their equilibrium conformation.
In the equation above, $k(c)$ represents a solute-dependent stiffness, which models the effects of the solute on the conformation of a polymer molecule. This function could be obtained from experiments measuring the persistence length of polymers at different solute concentrations or could be estimated from scaling laws. In the case of polyelectrolytes, the persistence length of a polymer depends on the local salt concentration and the function $k(c)$ can be estimated directly from theoretical considerations \cite{DOBRYNIN_2005}. The persistence length, $l_p$, of polyelectrolytes is proportional to the Debye-length, which scales as $c^{-1/2}$ with $c$ the number density of salt. For a flexible worm-like chain, the Kuhn length, $b$, is twice the persistence length \cite{rubinstein2003polymer}. It follows that the entropic elasticity of a polymer, which is equal to $3 k_\text{B}T/N b^2$ with $N$ the number of monomers, is a function of the local salt concentration. Using the scaling $l_p \propto c^{-1/2}$ and $b=2 \, l_p$, we can estimate the function $k(c)$ for the case of a polyelectrolyte in an ionic suspension as $k(c) \propto c$, where $c$ is the local salt concentration. 
We emphasize that the elastic energy given by Eq. \eqref{elast_energ} is a simple but rather crude approximation of the true elastic energy stored by a stretched polymer and it is expected to be valid only for small deformations. It is known that the choice of a linear spring leads to the divergence of the extensional viscosity at large extension rates \cite{larson2013constitutive}. Nevertheless, in this work, we focus on the effect of a solute-dependent stiffness and we leave the extension of Equation \eqref{elast_energ} to include finite spring extensibility or excluded volume effects \cite{Larson_2005} to future works. Throughout the rest of the paper, we drop the dependence of the dumbbell distribution, $\psi$, on $\boldsymbol{x}$, $t$ and $\boldsymbol{r}$ and the dependence of the solute distribution, $c$, on $\boldsymbol{x}$ and $t$ for clarity. 

\section{Derivation of the governing equations}
We begin the derivation of the governing equations by identifying the relevant conservation equations for the dumbbell and solute distribution.
The conservation of the number density of each species implies that the distribution of dumbbells must satisfy \cite{bird1987dynamics}
\begin{equation}\label{cont_eq_dumbb}
\frac{\partial \, \psi}{\partial t} + \frac{\partial}{\partial \boldsymbol{r}} \cdot (\dot{\boldsymbol{r}} \, \psi )+ \frac{\partial}{\partial \boldsymbol{x}} \cdot (\boldsymbol{w}_\text{d} \, \psi ) = 0 \, \, .
\end{equation} 
In Eq. \eqref{cont_eq_dumbb}, $\dot{\boldsymbol{r}}$ represents the rate of change of the end-to-end vector of the dumbbells and $\boldsymbol{w}_d$ is the velocity of the center of mass of the dumbbells. The conservation of the number density of the solute reads
\begin{equation}\label{cont_eq_sol}
\frac{\partial \, c}{\partial t}  + \frac{\partial}{\partial \boldsymbol{x}} \cdot (\boldsymbol{w} \, c ) =0 \, \, ,
\end{equation} 
where $\boldsymbol{w}$ is the velocity of the solute.
Mass conservation for an incompressible solvent implies:
\begin{equation}
\frac{\partial}{\partial \boldsymbol{x}} \cdot \boldsymbol{v} =0 \, \, ,
\end{equation}
where $\boldsymbol{v} $ is the velocity of the solvent.
We assume that the boundaries of the volume, $\delta \Omega$, are impermeable to the dumbbells, to the solute and to the solvent: 
\begin{equation}
\boldsymbol{w}_d \cdot \boldsymbol{n} =0; \, \, \, \, \, \,  \boldsymbol{w} \cdot \boldsymbol{n} =0   \, \, \, \, \, \,  \boldsymbol{v} \cdot \boldsymbol{n} =0   \, \, \, \, \, \, \text{on} \, \, \, \, \, \, \delta \Omega \, \, .
\end{equation}
Where $\boldsymbol{n}$ is the vector normal to the volume boundaries. Note that the framework developed here can be extended to consider adsorption and desorption of solute and dumbbells from the surfaces. Finally, the distribution of dumbbells goes to zero $\psi \rightarrow 0$ as the end-to-end vector goes to infinity $\boldsymbol{r} \rightarrow \infty $: there are no dumbbells that are infinitely stretched.

To derive the balance of momentum and the expressions for $\dot{\boldsymbol{r}}$, $\boldsymbol{w}_d$ and $\boldsymbol{w}$, we neglect inertial effects and we employ the Onsager's variational formalism \cite{doi2011onsager,arroyo2018onsager}. This framework allows us to identify the relevant couplings between the fluxes of solute, of dumbbells and of momentum directly from the principle of least dissipation. Briefly, the governing equations are obtained by minimizing the Rayleighian functional, constrained by the incompressibility condition, with respect to the process variables. In our case, the process variables are given by the velocity of the dumbbells, $\boldsymbol{w}_d$, by the velocity of the solute, $\boldsymbol{w}$, by the velocity of the solvent, $\boldsymbol{v}$, and by the rate of change of the end-to-end vector $\dot{\boldsymbol{r}}$. To constrain the Rayleighian to fulfill the incompressibility condition $\frac{\partial}{\partial \boldsymbol{x}} \cdot \boldsymbol{v}=0$, we include the pressure $p$ as Lagrange multiplier and we define the Lagrangian functional as 
\begin{equation}\label{Lagrangian_def}
\mathcal{L} = \mathcal{\dot{F}} + \mathcal{D} - \int_{\Omega} p \, \frac{\partial}{\partial \boldsymbol{x}} \cdot \boldsymbol{v} \, d\Omega \, \, ,
\end{equation}
In Eq. \eqref{Lagrangian_def}, $\mathcal{\dot{F}}$ and $\mathcal{D}$ are the rate of change of the free energy and the energy dissipation of the system, respectively. By minimizing $\mathcal{L}$ with respect to the process variables and with respect to the pressure we obtain the governing equations. Recently, Zhou and Doi \cite{Zhou_2018} showed that the Onsager's variational principle can be used to derive the Oldroyd-B constitutive model in the case of a homogeneous suspension of dumbbells.
Our approach bears some similarities to the GENERIC and bracket frameworks of nonequilibrium thermodynamics \cite{beris1994thermodynamics,ottinger2005beyond}, which have been used successfully to derive constitutive equations of complex fluids that couple different physics \cite{Mavrantzas_1992,V_zquez_Quesada_2009,Germann_2013,Germann_2016,H_tter_2020,Stephanou_2020}. While it is more transparent than the GENERIC formalism, our approach implicitly assumes that the momentum of all the fields are fast variables and relax quickly compared to the evolution of the system \cite{doi2013soft}.

\subsection{Rate of change of the free energy}
We begin by defining the free energy functional of the system, $\mathcal{F}$, as the sum of the contribution of the dumbbells and the solute %We identify the distribution of dumbbells, $\psi$, and the distribution of solute $c$ as state variables. 
\begin{equation}\label{free_energ}
\mathcal{F} = k_\text{B} T \int_\Omega c \log{c} \, d\Omega + k_\text{B} T \int_\Omega \int_{\boldsymbol{r}} \psi \log{\psi} \, d\Omega \, d\boldsymbol{r}+\int_\Omega \int_{\boldsymbol{r}} \psi U  \, d\Omega \, d\boldsymbol{r}  \, \, ,
\end{equation}
where $k_\text{B}$ is the Boltzmann's constant, $T$ is the absolute temperature and $U$ is the elastic energy, defined in Eq. \eqref{elast_energ}. The first two terms represent the solute and dumbbell free energy of mixing and the third term represents the elastic energy stored by the springs. Since the stiffness of the spring, $k(c)$, depends on the local solute concentration, the energy stored by the dumbbells also depends on $c$. The rate of change of the free energy $\mathcal{\dot{F}}$ is obtained by taking the time derivative of Eq. \eqref{free_energ}, see the Appendix \ref{appFreeEn} for details,
\begin{multline}\label{free_energ_time_deriv4}
\mathcal{\dot{F}} = k_\text{B} T \int_\Omega \frac{\partial c }{\partial \boldsymbol{x}}  \cdot \boldsymbol{w} \, d\Omega  +\int_\Omega \int_{\boldsymbol{r}} \left( k_\text{B} T \frac{\partial \psi}{\partial \boldsymbol{r}} +\psi \frac{\partial}{\partial \boldsymbol{r}}U \right) \cdot \dot{\boldsymbol{r}} \, d\Omega \, d\boldsymbol{r}+ \\
+ \int_\Omega \int_{\boldsymbol{r}} \left(k_\text{B} T \frac{\partial \psi}{\partial \boldsymbol{x}}+\psi \frac{\partial}{\partial \boldsymbol{x}}U \right) \cdot \boldsymbol{w}_\text{d}  \, d\Omega \, d\boldsymbol{r}+\int_\Omega \int_{\boldsymbol{r}} \frac{\partial}{\partial \boldsymbol{x}} \left( \frac{\partial U}{\partial c} \psi  \right) \cdot \left(  \boldsymbol{w} c \right) \, d\Omega \, d\boldsymbol{r} \, \, .
\end{multline}

\subsection{Energy dissipation}
We assume that the energy dissipation functional is comprised of three different contributions: (i) the friction due to the viscosity of the solvent, which is proportional to the solvent shear viscosity $\eta$ and to the rate of deformation tensor; (ii) the friction due to the relative motion of the solute molecules and the solvent, which is proportional to the friction coefficient of the solute $\xi$; and (iii) the friction between the dumbbells and the solvent, which is proportional to the friction coefficient of the dumbbell beads, $\xi_\text{d}$. For simplicity, we assume $\xi_\text{d}$ to be a constant and not to depend on the dumbbell extension \cite{Tanner_1975} or on the conformation of the nearby dumbbells as in the Giesekus model \cite{Giesekus_1982}. This is a simplification since we expect that changes in polymer conformation do impact the dissipation when moving relative to the solvent. Under these assumptions, the dissipation functional is given by:
\begin{multline}\label{dissipation}
\mathcal{D} =  \eta\int_{\Omega} \, \boldsymbol{D}:\boldsymbol{D} \, d\Omega +\frac{1}{2} \xi \int_{\Omega} \, c \left( \boldsymbol{w}- \boldsymbol{v}\right)^2 \, d\Omega +\\
+\xi_\text{d} \int_{\Omega} \int_{\boldsymbol{r}} \, \psi \left( \boldsymbol{w}_\text{d} - \boldsymbol{v}\right)^2 \, d\Omega \, d\boldsymbol{r} +\frac{1}{4} \xi_\text{d} \int_{\Omega} \int_{\boldsymbol{r}} \, \psi \left( \dot{\boldsymbol{r}} - \boldsymbol{r} \cdot \frac{\partial \boldsymbol{v}}{\partial \boldsymbol{x}} \right)^2 \, d\Omega \, d\boldsymbol{r}  \, \, .
\end{multline}
The first term in Eq. \eqref{dissipation} represents the dissipation due to the solvent rate of deformation,
\begin{equation} 
\boldsymbol{D}= \frac{1}{2} \left[\frac{\partial \boldsymbol{v}}{\partial \boldsymbol{x}} +\left(\frac{\partial \boldsymbol{v}}{\partial \boldsymbol{x}}\right )^T \right] \, \, ,
\end{equation} 
the second term represents the dissipation due to the relative motion of solute and solvent molecules. The last two terms of Eq. \eqref{dissipation} are the dissipation generated by the relative motion between the dumbbells and the solvent and can be derived from a Taylor expansion of the flow field around the dumbbell center of mass, as shown in the Appendix \ref{appA}. %Here we neglect the hydrodynamics interactions between multiple dumbbells and between the two dumbbell sphere \cite{Larson_2005}.

\subsection{Minimization of the Lagrangian}
To derive the governing equations, we substitute the rate of change of the free energy, given by Eq. \eqref{free_energ_time_deriv4}, and the dissipation functional, given by Eq. \eqref{dissipation}, in the definition of the Lagrangian, given by Eq. \eqref{Lagrangian_def}.
The minimization of $\mathcal{L}$ with respect to the solute velocity, $\boldsymbol{w}$, yields an expression for the solute flux:
\begin{equation}\label{solflux}
c \boldsymbol{w} = -\frac{k_\text{B} T}{\xi} \frac{\partial c}{\partial \boldsymbol{x}}  -\frac{1}{\xi} \int_{\boldsymbol{r}} \frac{\partial}{\partial \boldsymbol{x}} \left( \frac{\partial U}{\partial c} \psi  \right) c \, d\boldsymbol{r} + c \boldsymbol{v}\, \, .
\end{equation} 
The minimization of $\mathcal{L}$ with respect to the translational velocity of the dumbbells, $\boldsymbol{w}_d$, yields an expression for the spatial flux of dumbbells:
\begin{equation}\label{dumbflux1}
\psi \boldsymbol{w}_\text{d} = -\frac{k_\text{B} T}{2 \xi_\text{d}} \frac{\partial \psi}{\partial \boldsymbol{x}} -\frac{1}{2 \xi_\text{d}} \psi \frac{\partial U}{\partial \boldsymbol{x}} + \psi \boldsymbol{v} \, \, .
\end{equation} 
The minimization of $\mathcal{L}$ with respect to the rate of change of the end-to-end vector of the dumbbells, $\dot{\boldsymbol{r}}$, yields:
\begin{equation}\label{dumbflux2}
\psi \dot{\boldsymbol{r}} = -\frac{2 k_\text{B} T}{\xi_\text{d}} \frac{\partial \psi}{\partial \boldsymbol{r}} -\frac{2}{\xi_\text{d}} \psi \frac{\partial U}{\partial \boldsymbol{r}}  + \psi \boldsymbol{r} \cdot \frac{\partial \boldsymbol{v}}{\partial \boldsymbol{x}} \, \, .
\end{equation}
Finally, minimizing $\mathcal{L}$ with respect to the solvent velocity, $\boldsymbol{v}$, and using integration by parts yields the momentum balance: 
\begin{equation}\label{mombal}
-2 \eta \, \frac{\partial }{\partial \boldsymbol{x}} \cdot \boldsymbol{D} + \frac{\partial p}{\partial \boldsymbol{x}} + \xi  c (\boldsymbol{v}-\boldsymbol{w}) +2 \xi_\text{d}  \int_{\boldsymbol{r}} \psi (\boldsymbol{v}-\boldsymbol{w}_d) \, d\boldsymbol{r}+ \frac{\xi_\text{d}}{2} \frac{\partial }{\partial \boldsymbol{x}} \cdot \int_{\boldsymbol{r}} \boldsymbol{r} \psi \left(\dot{\boldsymbol{r}}-\boldsymbol{r} \cdot \frac{\partial \boldsymbol{v}}{\partial \boldsymbol{x}} \right) \, d\boldsymbol{r} = \boldsymbol{0} \,  \, .
\end{equation}

\subsection{Governing equations}
The governing equations are obtained by substituting the fluxes obtained in the previous section into the conservation equations Eqs. \eqref{cont_eq_dumbb} and \eqref{cont_eq_sol}. These equations represent one of the main results of this paper.
By substituting Eq. \eqref{solflux} into Eq. \eqref{cont_eq_sol} we obtain the transport equation of solute:
\begin{equation}\label{soltransp}
\frac{\partial c}{\partial t}  = \frac{\partial }{\partial \boldsymbol{x}} \cdot \left( \frac{k_\text{B} T}{\xi} \frac{\partial c}{\partial \boldsymbol{x}} +\frac{c}{2 \xi} \frac{\partial}{\partial \boldsymbol{x}} \left( \frac{d k(c)}{d c} \, \int_{\boldsymbol{r}} \psi \vert\boldsymbol{r}\vert^2  \, d\boldsymbol{r}\right) - c \boldsymbol{v} \right)\, \, ,
\end{equation} 
where we substituted the definition of elastic energy, $U$, as given by Eq. \eqref{elast_energ}. Besides the classic Fickian diffusion, with diffusion coefficient, $k_\text{B} T/\xi$, and the advection by the solvent flow, the right-hand side of Eq. \eqref{soltransp} displays one more contribution to the solute flux that is proportional to $ \frac{d k(c)}{d c}$. It represents a coupling between the distribution of dumbbells and the distribution of solute, which is introduced by the solute-dependent stiffness of the dumbbells. In the case of constant spring stiffness, the transport of solute reduces to the classic advection-diffusion equation. Since the integral of $\psi \, \vert\boldsymbol{r}\vert^2$ over $ \boldsymbol{r}$ represents the local average extension of the dumbbells, the new coupling term predicts that a gradient of dumbbell extension drives a flux of solute. As a consequence of this coupling, in the case of inhomogeneous flows where the stretching of polymers changes in space, we expect that the solute concentration is also inhomogeneous. The solute will migrate towards regions of higher or lower dumbbell extensions, depending on the sign of $ \frac{d k(c)}{d c}$. From a physical standpoint, the migration of solute is driven by the tendency of the solute to reduce the elastic energy stored by highly stretched dumbells. Therefore, if $ \frac{d k(c)}{d c}<0$, the solute migrates towards regions of large dumbbell stretch to make their spring softer and reduce the elastic energy. Conversely, if $ \frac{d k(c)}{d c}>0$, the solute migrates away from regions of large dumbbell extensions so to decrease the stiffness of the dumbbells in these regions. 

By substituting Eqs. \eqref{dumbflux1}-\eqref{dumbflux2} into the Eq. \eqref{cont_eq_dumbb} we obtain an equation for transport of dumbbells:
\begin{equation}\label{dumbtransp}
\frac{\partial \psi}{\partial t}  = \frac{\partial }{\partial \boldsymbol{r}} \cdot \left(\frac{2 k_\text{B} T}{\xi_\text{d}} \frac{\partial \psi}{\partial \boldsymbol{r}} +\frac{2}{\xi_\text{d}} k(c) \boldsymbol{r} \psi - \psi \boldsymbol{r} \cdot \frac{\partial }{\partial \boldsymbol{x}} \boldsymbol{v} \right)+\frac{\partial }{\partial \boldsymbol{x}} \cdot \left(\frac{k_\text{B} T}{2 \xi_\text{d}} \frac{\partial \psi}{\partial \boldsymbol{x}} +\frac{1}{4 \xi_\text{d}} \frac{d k(c)}{d c} \vert\boldsymbol{r}\vert^2 \psi \frac{\partial c}{\partial \boldsymbol{x}} - \psi \boldsymbol{v} \right)\, \, .
\end{equation}
The first bracket on the right-hand side of Eq. \eqref{dumbtransp} represents the flux in the end-to-end space, while the second term on the right-hand side represents the spatial flux of dumbbells.
The flux of dumbbells in the end-to-end vector space, $\boldsymbol{r}$, is composed of three terms: the first one is a Brownian relaxation term, the second one is the effect of the restoring force due to the spring and the third is the stretching due to straining solvent flows. This flux is  similar to that derived in other dumbbell kinetic models \cite{Larson_2005}, with the exception that here the spring stiffness depends on the local solute concentration. As a consequence, the relaxation time of the end-to-end vector, which is proportional to $\xi_\text{d}/ k(c)$, can change in space. The spatial flux of dumbbells has three contributions: a Fickian diffusion term with an effective diffusion coefficient given by $k_\text{B}T/2 \, \xi_\text{d}$, a contribution originating from the coupling with the solute field, and a contribution due to the advection by the solvent flow. The second contribution, proportional to $ \frac{d k(c)}{d c}$, is novel and it predicts that a gradient of solute drives a spatial flux of dumbbells. The flux arising from gradients of solute is proportional to $\psi \vert\boldsymbol{r}\vert^2$, which represents the local extension of the dumbbells. The more stretched are the dumbbells the larger is their spatial flux resulting from a gradient of solute. The direction of the dumbbell flux depends on the sign of $ \frac{d k(c)}{d c}$. The physical mechanism driving the migration of dumbbells is similar to that of the solute. In the case of an inhomogeneous distribution of solute, the dumbbells migrate towards regions of higher or lower concentrations of solute to reduce the elastic energy stored by the spring. We emphasize that this mechanism of polymer migration is fundamentally different from the well-known mechanisms driven by curved streamlines \cite{Aubert_1980,Aubert_1980_1} or by stress gradients \cite{Helfand_1989,Doi_1992,Cromer_2013,Tsouka_2014}.

By substituting Eqs. \eqref{dumbflux1}-\eqref{dumbflux2} into the momentum balance, given by Eq. \eqref{cont_eq_dumbb}, we obtain:
\begin{equation}\label{mombal1}
\frac{\partial }{\partial \boldsymbol{x}} \cdot \left(2 \eta \, \boldsymbol{D} - P \boldsymbol{I} +  k(c) \int_{\boldsymbol{r}} \boldsymbol{r}\boldsymbol{r} \psi \, d\boldsymbol{r} -k_BT n_\text{d} \boldsymbol{I} \right) = \boldsymbol{0}\,  \, ,
\end{equation}
where $\boldsymbol{r} \boldsymbol{r}$ denotes the dyadic product and $\boldsymbol{I}$ denotes the identity tensor.
In Eq. \eqref{mombal1}, we have defined a modified pressure, $P$, which is the sum of the hydrodynamic pressure and of the osmotic pressure:
\begin{equation}
P = p + k_\text{B}T \left( c +n_\text{d} \right) + \frac{1}{2} \left(c \frac{d k(c)}{d c} \int_{\boldsymbol{r}}  \psi \, \vert\boldsymbol{r}\vert^2 \, d\boldsymbol{r} \right)\, \, ,
\end{equation}
where $n_\text{d}$ is the local number density of dumbbells defined in Eq. \eqref{dumbb_conc}. The stress generated by the elasticity of the dumbells is proportional to the spring $k(c)$, which can vary in space. %The relative motion between the dumbbells and the solvent and between the solute and the solvent introduces body forces, which are collected in the right hand side of Eq. \eqref{mombal1}.
Finally, by maximizing the Lagrangian, given by Eq. \eqref{Lagrangian_def}, with respect to the pressure, $p$, we obtain the continuity equation
\begin{equation}\label{cont_eq_solv}
\frac{\partial }{\partial \boldsymbol{x}} \cdot \boldsymbol{v}=0 \, \, .
\end{equation}

By neglecting the dependence of the spring stiffness on the solute concentration the transport of solute decouples from the transport of dumbbells. If we also neglect the spatial variations of the dumbbell concentration, then Eqs. \eqref{dumbtransp} and Eq. \eqref{mombal1} reduce to the Stokes equation with the stress tensor given by the Oldroyd-B constitutive model \cite{larson2013constitutive}. This can be shown by multiplying Eq. \eqref{dumbtransp} by $\boldsymbol{r}\boldsymbol{r}$ and integrating over the end-to-end vector space. Interestingly, our results show that naively using an Oldroyd-B model with solute-dependent rheological properties would lead to the wrong set of equations because it neglects the couplings between the spatial gradients of solute and dumbbell distribution that we identified in Eqs. \eqref{soltransp}-\eqref{dumbtransp} . 
Once the boundary conditions and the initial conditions are specified, the set of Eqs. \eqref{soltransp}-\eqref{cont_eq_solv} yield a closed problem for the evolution of the fields $\boldsymbol{v}$, $c$, $\psi$ and $P$. 

\section{Equilibrium phase separation of a homogeneous suspension}\label{equiPS}
The well-established Flory-Huggins theory shows that changing the solvent quality can lead to phase separation \cite{rubinstein2003polymer}. Since changing the solute concentration can be interpreted as a change in solvent quality, it is interesting to ask if the simple model that we propose also predicts an equilibrium phase separation. To do so, we study the stability of a homogeneous suspension of solute and dumbbells at equilibrium using linear stability analysis. We assume small perturbations to the equilibrium concentration of solute, $c_\text{eq}$ and to the equilibrium dumbbell distribution $\psi_\text{eq} $:
\begin{equation}\label{perturb}
c= c_\text{eq}+c'; \, \, \, \psi = \psi_\text{eq} + \psi' \, \, .
\end{equation}
The equilibrium distribution of dumbbells is given by the Boltzmann distribution:
\begin{equation}
\psi_\text{eq} =  \frac{n_\text{d,eq}}{2 \sqrt{2}}\left(\frac{k_\text{eq}}{\pi \, k_\text{B}T}\right)^{3/2} \, \exp{\left(-\frac{\vert\boldsymbol{r}\vert^2 }{2} \, \frac{k_\text{eq}}{k_\text{B}T}\right)}\, \, ,
\end{equation}
where $n_\text{d,eq}$ is the number density of dumbbells and $k_\text{eq}$ denotes the stiffness of the dumbbell spring at equilibrium: it is a shorthand notation for $k(c_\text{eq})$. It is straightforward to show that the integral of $\psi_\text{eq} $ over the end-to-end vector space gives $n_\text{d,eq}$. 

By substituting the perturbations, given by Eqs. \eqref{perturb}, into the solute transport equation, given by Eq. \eqref{soltransp}, and neglecting the nonlinear terms, we obtain:
\begin{multline}\label{soltransp_pert}
\frac{\partial c'}{\partial t}  = \frac{\partial }{\partial \boldsymbol{x}} \cdot \Bigg( \frac{k_\text{B} T}{\xi} \frac{\partial c'}{\partial \boldsymbol{x}} +\frac{c_\text{eq}}{2 \xi} \frac{d k(c)}{d c}\Bigr|_{c_\text{eq}} \, \frac{\partial}{\partial \boldsymbol{x}} \left(  \int_{\boldsymbol{r}} \psi'  \, \vert\boldsymbol{r}\vert^2   \, d\boldsymbol{r}\right) +\frac{c_\text{eq}}{2 \xi} \frac{d^2 k(c)}{d c^2}\Bigr|_{c_\text{eq}} \, \frac{\partial}{\partial \boldsymbol{x}} \left( c' \, \int_{\boldsymbol{r}} \psi_\text{eq} \, \vert\boldsymbol{r}\vert^2  \, d\boldsymbol{r}\Bigg) \right)\, \, ,
\end{multline} 
where we have Taylor expanded $\frac{d k(c)}{d c}$ around $c=c_\text{eq}$. We can further simplify Eq. \eqref{soltransp_pert} and carry out the last integral in the right-hand side: 
\begin{equation}
\int_{\boldsymbol{r}} \psi_\text{eq} \, \vert\boldsymbol{r}\vert^2  \, d\boldsymbol{r} =  \frac{3 \,n_\text{d,eq} \,  k_\text{B} T}{k_\text{eq}} \, \, ,
\end{equation}
and we define the perturbation to the equilibrium stretching of the dumbbells as $S'$:
\begin{equation}
S'=\int_{\boldsymbol{r}} \psi' \, \vert\boldsymbol{r}\vert^2 \, d\boldsymbol{r}  \, \, .
\end{equation}
We then rewrite Eq. \eqref{soltransp_pert} as:
\begin{equation}\label{soltransp_pert1}
\frac{\partial c'}{\partial t}  = \frac{\partial }{\partial \boldsymbol{x}} \cdot \left( \frac{k_\text{B} T}{\xi} \frac{\partial c'}{\partial \boldsymbol{x}} +\frac{c_\text{eq}}{2 \xi} \frac{d k(c)}{d c}\Bigr|_{c_\text{eq}} \, \frac{\partial S'}{\partial \boldsymbol{x}}  +\frac{3 c_\text{eq}}{2 \xi} \frac{n_\text{d,eq} k_\text{B} T}{k_\text{eq}} \frac{d^2 k(c)}{d c^2}\Bigr|_{c_\text{eq}} \, \frac{\partial}{\partial \boldsymbol{x}} c'  \right)\, \, .
\end{equation} 
Next, we substitute Eq. \eqref{perturb} into Eq. \eqref{dumbtransp} and we neglect the nonlinear terms: 
\begin{multline}\label{dumbtransp_pert}
\frac{\partial \psi'}{\partial t}  = \frac{\partial }{\partial \boldsymbol{r}} \cdot \left(\frac{2 k_\text{B} T}{\xi_\text{d}} \frac{\partial \psi'}{\partial \boldsymbol{r}} +\frac{2}{\xi_\text{d}} k_\text{eq} \boldsymbol{r} \psi' +\frac{2}{\xi_\text{d}} \frac{d k(c)}{d c}\Bigr|_{c_\text{eq}} \boldsymbol{r} \psi_{eq} c'  \right)+\\
+\frac{\partial }{\partial \boldsymbol{x}} \cdot \left(\frac{k_\text{B} T}{2 \xi_\text{d}} \frac{\partial \psi'}{\partial \boldsymbol{x}} +\frac{1}{4 \xi_\text{d}} \frac{d k(c)}{d c}\Bigr|_{c_\text{eq}} \, \vert\boldsymbol{r}\vert^2 \,\psi_\text{eq} \frac{\partial c'}{\partial \boldsymbol{x}} \right)\, \, .
\end{multline}
The transport of solute, given by Eq. \eqref{soltransp_pert1}, shows that the evolution of the perturbations to the solute concentration depends only on the dumbbell stretch $S'$ and not on the higher moments of the dumbbell distribution. To obtain an equation for $S'$, we multiply Eq. \eqref{dumbtransp_pert} by $\boldsymbol{r}^2$ and we integrate over the end-to-end vector space using integration by parts:
\begin{multline}\label{stretch_pert}
\frac{\partial S'}{\partial t}  = \frac{4}{\xi_\text{d}}\left(3 k_\text{B} T n_\text{d}' - k_\text{eq} S' -3 \frac{d k(c)}{d c}\Bigr|_{c_\text{eq}} \frac{n_\text{d,eq} k_\text{B} T}{k_\text{eq}} c'  \right)+\\
+\frac{\partial }{\partial \boldsymbol{x}} \cdot \left(\frac{k_\text{B} T}{2 \xi_\text{d}} \frac{\partial S'}{\partial \boldsymbol{x}} +\frac{1}{4 \xi_\text{d}} \frac{d k(c)}{d c}\Bigr|_{c_\text{eq}} \int_{\boldsymbol{r}}  \, \vert\boldsymbol{r}\vert^4 \, \psi_\text{eq} \, d\boldsymbol{r} \frac{\partial c'}{\partial \boldsymbol{x}} \right)\, \, .
\end{multline}
Where $n_\text{d}'$ is the perturbation to the local number density of dumbbells defines as $n_\text{d}'=\int_{\boldsymbol{r}} \psi' d\boldsymbol{r}$.
The last integral in the right hand side of Eq. \eqref{stretch_pert} can be computed to give:
\begin{equation}
\int_{\boldsymbol{r}}  \, \vert\boldsymbol{r}\vert^4 \, \psi_\text{eq} \, d\boldsymbol{r} = 15 n_\text{d,eq} \left(\frac{k_\text{B} T}{k_\text{eq}}\right)^2 \, \, ,
\end{equation}
which substituted in Eq. \eqref{stretch_pert} yields: 
\begin{multline}\label{stretch_pert1}
\frac{\partial S'}{\partial t}  = \frac{4}{\xi_\text{d}}\left(3 k_\text{B} T n_\text{d}' - k_\text{eq} S' -3 \frac{d k(c)}{d c}\Bigr|_{c_\text{eq}} \frac{n_\text{d,eq} k_\text{B} T}{k_\text{eq}} c'  \right)+\\
+\frac{\partial }{\partial \boldsymbol{x}} \cdot \left(\frac{k_\text{B} T}{2 \xi_\text{d}} \frac{\partial S'}{\partial \boldsymbol{x}} +\frac{15}{4 \xi_\text{d}} \frac{d k(c)}{d c}\Bigr|_{c_\text{eq}} n_\text{d,eq} \left(\frac{k_\text{B} T}{k_\text{eq}}\right)^2 \frac{\partial c'}{\partial \boldsymbol{x}} \right)\, \, .
\end{multline}
To obtain an equation for $n_\text{d}'$, we integrate Eq. \eqref{dumbtransp_pert} over the end-to-end vector space using integration by parts:
\begin{equation}\label{dumbconc_pert}
\frac{\partial n_\text{d}'}{\partial t}  = \frac{\partial }{\partial \boldsymbol{x}} \cdot \left(\frac{k_\text{B} T}{2 \xi_\text{d}} \frac{\partial n_\text{d}'}{\partial \boldsymbol{x}} +\frac{3}{4 \xi_\text{d}} \frac{d k(c)}{d c}\Bigr|_{c_\text{eq}} \frac{n_\text{d,eq} \, k_\text{B} T}{k_\text{eq}} \frac{\partial c'}{\partial \boldsymbol{x}} \right)\, \, .
\end{equation}
Finally, it can be shown that the distribution of the end-to-end vector remains isotropic during the evolution of a perturbation. Eq. \eqref{mombal1} shows that an isotropic end-to-end vector distribution only modifies the pressure but it does not generate a velocity field. As a consequence, we do not need to consider perturbations of the velocity field.

Eqs. \eqref{soltransp_pert1}, \eqref{stretch_pert1} and \eqref{dumbconc_pert} represent a linear system of partial differential equations that govern the evolution of small perturbations to the homogeneous state, which can be solved using standard methods. Here we assume an infinite domain and we consider the following ansatz for the perturbations $c'$, $n_\text{d}'$ and $S'$:
\begin{align}\label{ansatz}
S'= S_0' \exp(\lambda t + i \boldsymbol{x} \cdot \mathbf{k}) \, \, ,\\ 
n_\text{d}'=n_{d,0}' \exp(\lambda t + i \boldsymbol{x} \cdot \mathbf{k}) \, \, , \\
c'= c_0' \exp(\lambda t + i \boldsymbol{x} \cdot \mathbf{k}) \, \, .
\end{align}
Where $i$ is the imaginary unit, $\mathbf{k}$ is the wavevector of the perturbation, $\lambda$ is the growth rate and $S_0'$, $n_{d,0}'$ and $c_0'$ are the initial values of the perturbations. Since the equations considered in the stability analysis are linear, the initial values of the perturbations are arbitrary and do not play any role in the stability of the equilibrium state. By inserting the ansatz, given by Eq. \eqref{ansatz}, into Eqs. \eqref{soltransp_pert1}, \eqref{stretch_pert1} and \eqref{dumbconc_pert} we obtain an eigenvalue problem for the growth rate $\lambda$, which has to be solved numerically. Once the parameters are specified, the numerical solution yields the eigenvalues as a function of the wavevector. 

For all the parameters tested we find three real eigenvalues. The homogeneous state is stable when all the three eigenvalues are negative but it is unstable when one or more eigenvalues are positive. We find that one eigenvalue can be positive, depending on the parameters chosen. Since in this section we examine thermodynamic phase separation, the stability of the homogeneous state depends on three dimensionless numbers that do not contain any transport coefficient: the number density of dumbbells relative to that of the solute $\alpha=n_\text{d,eq}/c_\text{eq}$, the dimensionless slope of the dumbbell spring stiffness evaluated at the homogeneous state $\beta=(c_\text{eq}/k_\text{eq}) d k(c)/d c\Bigr|_{c=c_\text{eq}}$, and the dimensionless curvature of the dumbbell spring stiffness evaluated at the homogeneous state $\gamma=(c_\text{eq}^2/k_\text{eq}) d^2 k(c)/d c^2\Bigr|_{c=c_\text{eq}}$. To discriminate between stable and unstable homogeneous states, we look for the set of parameters for which at least one eigenvalue has a positive real part at any wavevector. 

In appendix \ref{app_pert_eq} we show that the stability of the homogeneous phase can also be studied considering perturbations of the free energy, given by Eq. \eqref{free_energ}. Briefly, we look at how the total free energy of the system changes when the fields $c$ and $\psi$ are perturbed with respect to their equilibrium values. If the perturbations reduce the total free energy, then the homogeneous system is unstable and phase separates. Using this approach, we find that the equilibrium homogeneous state is unstable if 
\begin{equation}
\frac{4}{15} \,  \alpha\, \beta^2 -6  \alpha \, \gamma -1>0 \, \, .
\end{equation} 
It follows that, the separation between stable and unstable regions in the parameter space is determined by the equation: 
\begin{equation}\label{Thermo_stab}
\alpha =  \frac{4} {15 \, \beta^2 -6  \, \gamma  }\, \, .
\end{equation} 
This analytical expression can be used to verify the results obtained from the numerical solution of the eigenvalue problem. Interestingly, Eq. \eqref{Thermo_stab} also shows that the homogeneous equilibrium state is always stable if $ \gamma > 15/6 \, \beta^2$. 

\begin{figure}[h!]
\centering
\includegraphics[width=0.75\textwidth]{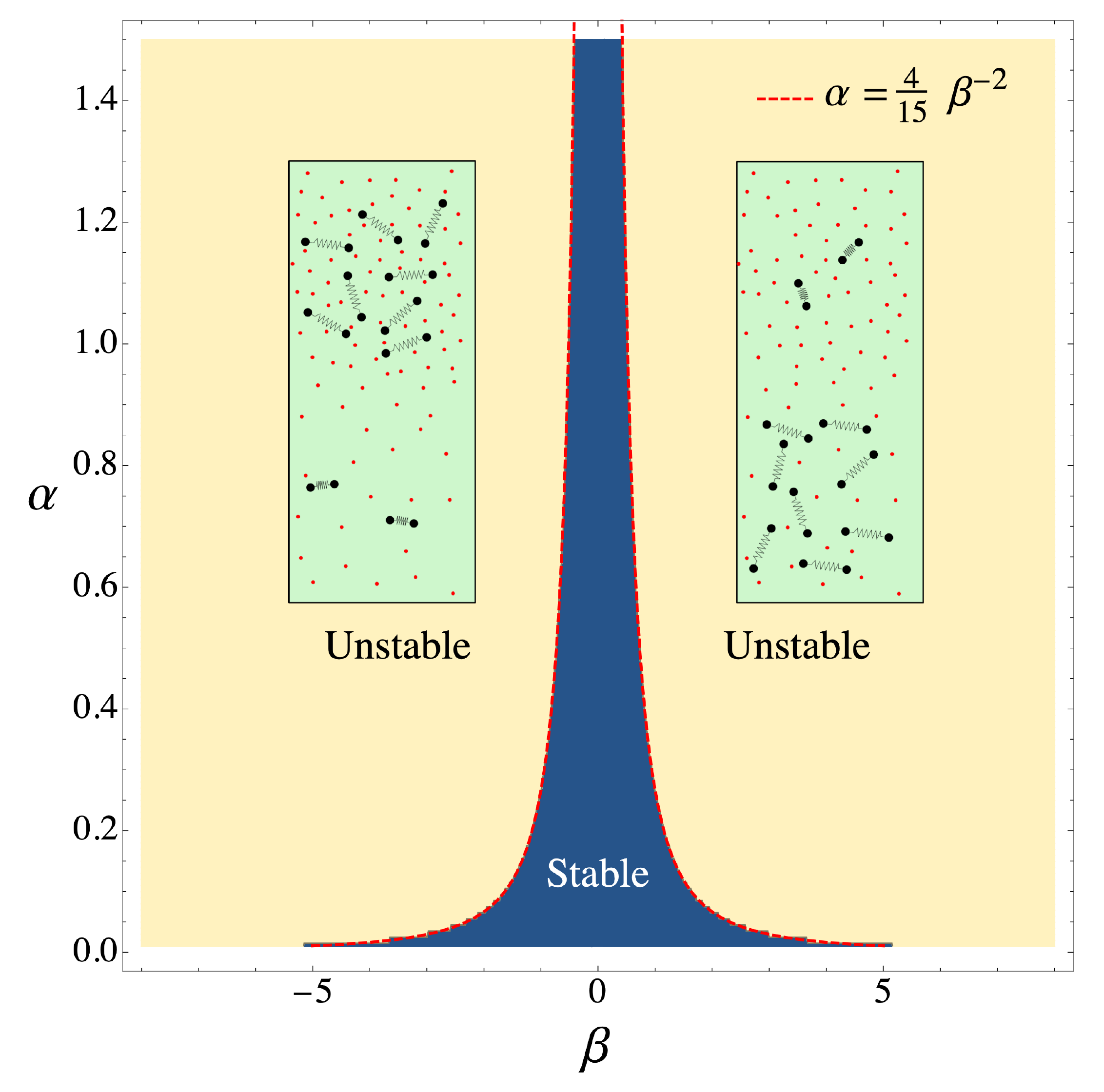}
\caption{Stability diagram of a homogeneous suspension of solute and dumbbells as a function of the ratio of their concentrations $\alpha$ and the dimensionless derivative of the spring stiffness with respect to the solute concentration $\beta$. The stability diagram is even with respect to changes of $\beta$. The inset shows the distribution of solute and dumbbells towards which the homogeneous evolves. The remaining dimensionless number is fixed zero, $\gamma=0$. The blue and yellow regions are obtained by solving numerically the eigenvalue problem while the red dashed line shows the prediction of Eq. \eqref{Thermo_stab}, which is obtained from the perturbation of the free energy. }
\label{fig2}
\end{figure}

In Figure \ref{fig2}, we report the stability diagram of a homogeneous distribution of solute and dumbbells for the case of $\gamma=0$. We plot the stable and the unstable regions as a function of the two remaining dimensionless numbers $\alpha$, and $\beta$. The regions of stability obtained through the numerical evaluation of the eigenvalues are shown in blue and yellow and they match those predicted by Eq. \eqref{Thermo_stab}. For a given ratio of dumbbell to solute number density, $\alpha$, there exist two values of $\beta$ beyond which the homogeneous distribution becomes unstable. As predicted by Eq. \eqref{Thermo_stab}, we find that the boundary separating the stable and unstable regions is symmetric with respect to the axis $\beta=0$. It follows that the stability of the homogeneous phase depends on the magnitude of $\beta$ but not on its sign.  From a physical standpoint this means that it is not relevant if the solute makes the spring of the dumbbells stiffer or softer but only how rapidly the spring stiffness changes by changing the solute number density. The sign of $\beta$ determines how dumbbells and of solute are distributed at the end of the phase separation. In the case of $\beta>0$, the dumbbells accumulate in regions of low solute concentration. Conversely, in the case of $\beta<0$, the dumbbells and the solute accumulate in the same regions. These distributions are schematically depicted as an inset in Figure \ref{fig2}.

\begin{figure}[h!]
\centering
\includegraphics[width=1.0\textwidth]{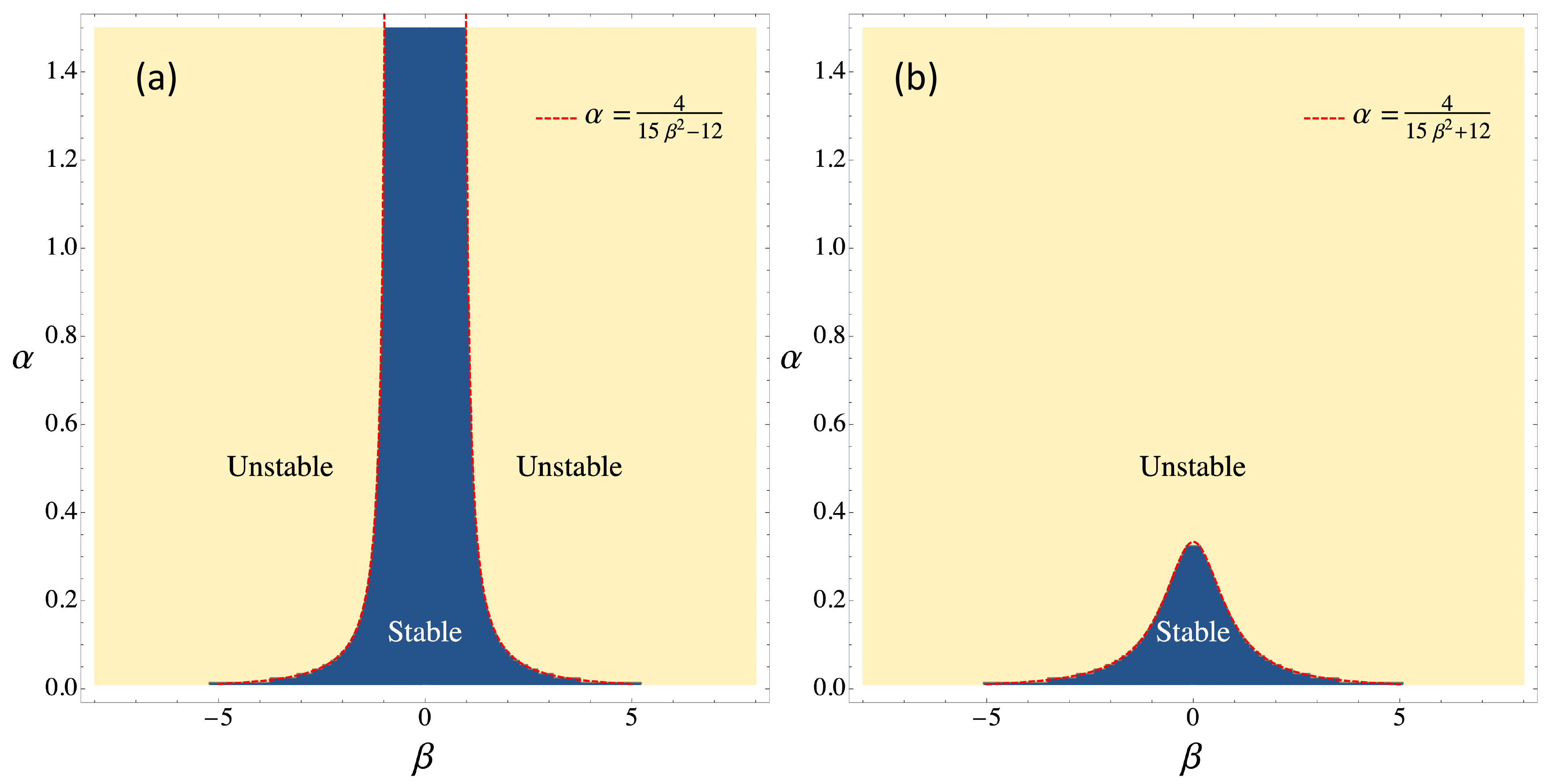}
\caption{Stability diagram of a homogeneous suspension of solute and dumbbells as a function of the ratio of their concentrations $\alpha$ and the dimensionless derivative of the spring stiffness with respect to the solute concentration $\beta$. In the panel (a) $\gamma=2$ while in the panel (b) $\gamma=-2$. The blue and yellow regions are obtained by solving numerically the eigenvalue problem while the red dashed line shows the prediction of Eq. \eqref{Thermo_stab}, which is obtained from the perturbation of the free energy. }
\label{fig3}
\end{figure}

In Figure \ref{fig3}, we report the effects of a nonzero second derivative of the spring stiffness with respect to the dimensionless solute concentration $\gamma$. In Figure \ref{fig3}(a), we find that a positive curvature $\gamma>0$ tends to stabilize the homogeneous distribution. Compared to the case shown in Figure \ref{fig2},  the boundary of stability is pushed towards larger values of $\beta$. This means  that a steeper slope of $k(c)$ is required to make the homogeneous distribution unstable. Conversely, Figure \ref{fig3}(b) shows that a negative curvature $\gamma<0$ tends to destabilize the homogeneous distribution. In this case, the homogeneous distribution can become unstable even in the case of $\beta=0$.  In the same plots, we show that the numerical solution of the eigenvalue problem matches perfectly the stability boundaries predicted by Eq. \eqref{Thermo_stab}, which is shown as a red dashed line. %Finally, repeating the linear stability analysis with we find that the dimensionless number $\Lambda \, c_\text{eq} k_\text{eq} (k_\text{B}T)^{-2}$ has a limited impact on the stability boundaries of a homogeneous distribution. Considering $\Lambda \, c_\text{eq} k_\text{eq} (k_\text{B}T)^{-2} \approx 1$ hardly changes the stability boundaries shown in Figs. \ref{fig2}-\ref{fig3}. 
In the present work, we do not consider any mechanism that stabilizes the phase boundaries at longer stages of the phase separation. If one is interested in the long-time and nonlinear behavior of phase separating systems, a term that penalizes the gradients of dumbbell concentration should be included in the formulation of the free energy, given by Eq. \eqref{free_energ}.

The spontaneous phase separation of a homogeneous distribution of solute and dumbbells can be understood, qualitatively, as an entropy-driven phase separation. The number of microstates that each dumbbell has access to depends on the stiffness of its spring. Dumbbells with softer springs access more microstates than dumbbells with a stiffer spring, therefore their entropy is larger. In the case of stiffness that changes significantly with the solute concentration, the dumbbells aggregate in regions where their spring is softer to maximize the total entropy. A phase-separated state, which apparently has a larger degree of spatial order, has, in fact, a larger total entropy than a homogeneous state. 
Finally, the results displayed in Figures \ref{fig2}-\ref{fig3} show that the model developed here, despite its simplicity, displays a similar phase transition to that predicted by the Flory-Huggins theory. 

\section{A closure approximation for the conformation tensor in the case of generic flows}\label{closureapproximation}
One of the drawbacks of Eq. \eqref{dumbtransp} is that it has to be solved over the space domain $\boldsymbol{x}$ and the end-to-end vector space $\boldsymbol{r}$. This was not an obstacle to performing the linear stability analysis  in the previous section but it complicates significantly numerical simulations. In this section, we discuss how this problem can be mitigated. We begin by realizing that the transport of solute, Eq. \eqref{soltransp}, and the momentum balance, Eq. \eqref{mombal1}, do not require the knowledge of the entire distribution of the end-to-end vectors but only of its second moment. The second moment of the end-to-end distribution has been often called the conformation tensor of the dumbbells and we define it as:
\begin{equation}
\boldsymbol{C} =  \int_{\boldsymbol{r}}  \psi \, \boldsymbol{r} \boldsymbol{r} \, d\boldsymbol{r} \, \, .
\end{equation}
The conformation tensor is a symmetric tensor and contains information about the local stretch of the dumbbells and the principal directions of stretch. For dumbbells that are stretched equally along all the directions, the conformation tensor is proportional to the identity tensor, $\boldsymbol{I}$. Note that, according to our definition, the conformation tensor also contains information about the local number density of dumbbells.
Another related quantity appearing in Eq. \eqref{soltransp} and in Eq. \eqref{mombal1} is the trace of the conformation tensor:
\begin{equation}
Tr(\boldsymbol{C}) =  \int_{\boldsymbol{r}}  \psi \, \vert\boldsymbol{r}\vert^2 \, d\boldsymbol{r} \, \, ,
\end{equation}
which is proportional to the local average stretch of the dumbbells.

In principle, one needs to solve the transport equation of the dumbbells, given by Eq. \eqref{dumbtransp}, over both the space domain $\boldsymbol{x}$ and the end-to-end vector space $\boldsymbol{r}$ and then integrate it to compute the conformation tensor and its trace. This requires the solution of a set of partial differential equations over a six-dimensional space, which is notoriously complicated, especially using numerical simulations. Instead, inspired by previous works on polymer suspensions \cite{bird1987dynamics,Larson_2005}, we derive an equation for the conformation tensor, $\boldsymbol{C} $. By multiplying Eq. \eqref{dumbtransp} with $\boldsymbol{r}\boldsymbol{r}$ and then integrating it over the end-to-end vector space, $\boldsymbol{r}$, we obtain: 
\begin{multline}\label{conftenstransp}
\int_{\boldsymbol{r}} \boldsymbol{r} \boldsymbol{r} \frac{\partial \psi}{\partial t} \, d\boldsymbol{r} = \int_{\boldsymbol{r}} \boldsymbol{r} \boldsymbol{r} \frac{\partial }{\partial \boldsymbol{r}} \cdot \left(\frac{2 k_\text{B} T}{\xi_\text{d}} \frac{\partial \psi}{\partial \boldsymbol{r}} +\frac{2}{\xi_\text{d}} k(c) \boldsymbol{r} \psi - \psi \boldsymbol{r} \cdot \frac{\partial }{\partial \boldsymbol{x}} \boldsymbol{v} \right) \, d\boldsymbol{r} +\\
 + \int_{\boldsymbol{r}} \boldsymbol{r} \boldsymbol{r} \frac{\partial }{\partial \boldsymbol{x}} \cdot \left(\frac{k_\text{B} T}{2 \xi_\text{d}} \frac{\partial \psi}{\partial \boldsymbol{x}} +\frac{1}{4 \xi_\text{d}} \frac{d k(c)}{d c} \, \vert\boldsymbol{r}\vert^2 \, \psi \frac{\partial c}{\partial \boldsymbol{x}} - \psi \boldsymbol{v} \right) \, d\boldsymbol{r}\, \, .
\end{multline}
We apply integration by parts in the first integral of the right hand side of Eq. \eqref{conftenstransp}. After some manipulation, we obtain:
\begin{multline}\label{conftenstransp1}
\frac{\partial \boldsymbol{C}}{\partial t} + \boldsymbol{v} \cdot \frac{\partial }{\partial \boldsymbol{x}} \boldsymbol{C}  -\left( \frac{\partial }{\partial \boldsymbol{x}}\boldsymbol{v}\right)^T \cdot \boldsymbol{C}-\boldsymbol{C} \cdot \frac{\partial }{\partial \boldsymbol{x}}\boldsymbol{v} = -\frac{4}{\xi_\text{d}}\left( k(c) \boldsymbol{C}-k_\text{B}T n_\text{d} \boldsymbol{I}\right) +\\
 + \frac{k_\text{B} T}{2 \xi_\text{d}} \left(\frac{\partial }{\partial \boldsymbol{x}}\right)^2 \boldsymbol{C} + \frac{1}{4 \xi_\text{d}} \frac{\partial }{\partial \boldsymbol{x}} \cdot \int_{\boldsymbol{r}} \boldsymbol{r} \boldsymbol{r} \left( \frac{d k(c)}{d c} \, \vert\boldsymbol{r}\vert^2 \, \psi \frac{\partial c}{\partial \boldsymbol{x}} \right) \, d\boldsymbol{r}\, \, .
\end{multline}
In Eq. \eqref{conftenstransp1}, the left-hand side is the upper convected time derivative of the conformation tensor and $n_\text{d}$ represents the local number density of dumbbells. In the last term on the right-hand side, the scalar product is carried out between the two spatial derivatives. Eq. \eqref{conftenstransp1} shows that the conformation tensor diffuses with a diffusion coefficient proportional to $k_\text{B} T/2 \xi_\text{d}$. The last term in the right-hand side of Eq. \eqref{conftenstransp1} shows that gradients of solute also contribute to a spatial flux of the conformation tensor. Their contribution is proportional to higher moments of the end-to-end vector distribution. To obtain this term, we can multiply Eq. \eqref{dumbtransp} by $\vert\boldsymbol{r}\vert^2 \, \boldsymbol{r}\boldsymbol{r}$ and then integrate over the end-to-end vector space. However, this would involve again higher moments of the distribution. Repeating this procedure iteratively leads to an infinite relation between higher and higher moments of $\psi$, which is not practical. A similar problem is encountered in microscopic models for finite-extensive springs or rigid rods  \cite{doi1988theory,Larson_2005}, and has been addressed by relating higher-order moments to products of lower-order ones. Inspired by these early works, we propose the following closure approximation:
\begin{equation}\label{closure_app}
\int_{\boldsymbol{r}} \boldsymbol{r} \boldsymbol{r} \, \vert\boldsymbol{r}\vert^2 \, \psi \, d\boldsymbol{r} \approx \frac{5 Tr(\boldsymbol{C})}{3\Psi} \boldsymbol{C} \, \, , 
\end{equation} 
where the factor $5/3$ is included so that the right and the left-hand side are equal at equilibrium. The proposed closure approximation can be interpreted as pre-averaging the term of Eq. \eqref{dumbtransp} proportional to $\vert\boldsymbol{r}\vert^2 \, \psi $. 

By inserting this approximation in Eq. \eqref{conftenstransp1}, we obtain:
\begin{equation}\label{conftenstransp2}
\overset{\nabla}{\boldsymbol{C}} = -\frac{4}{\xi_\text{d}}\left( k(c) \boldsymbol{C}-k_\text{B}T \Psi \boldsymbol{I}\right) + \frac{k_\text{B} T}{2 \xi_\text{d}} \left(\frac{\partial }{\partial \boldsymbol{x}}\right)^2 \boldsymbol{C} + \frac{5}{12 \xi_\text{d}} \frac{\partial }{\partial \boldsymbol{x}} \cdot \left( \frac{d k(c)}{d c} \frac{\partial c}{\partial \boldsymbol{x}} \frac{Tr(\boldsymbol{C})}{\Psi} \boldsymbol{C}\right) \, \, ,
\end{equation}
where 
\begin{equation}
\overset{\nabla}{\boldsymbol{C}}=\frac{\partial \boldsymbol{C}}{\partial t} + \boldsymbol{v} \cdot \frac{\partial }{\partial \boldsymbol{x}} \boldsymbol{C} -\left( \frac{\partial }{\partial \boldsymbol{x}}\boldsymbol{v}\right)^T \cdot \boldsymbol{C}-\boldsymbol{C} \cdot \frac{\partial }{\partial \boldsymbol{x}}\boldsymbol{v} \, \, ,
\end{equation}
denotes the upper convected derivative of the tensor $\boldsymbol{C}$.
Note that, in contrast to previous works \cite{Larson_2005}, the closure approximation is expected to have limited effects on the evolution of the conformation tensor. Indeed, as we discuss more in depth at the end of section \ref{Dimensionless}, the relaxation of the dumbbells is usually much faster than the time required to diffuse over macroscopic lengths. In this case, the last two terms on the right-hand side of Eq. \eqref{conftenstransp2} have a limited impact on the dynamics of the conformation tensor.

The right hand side of Eq. \eqref{conftenstransp2}, contains the local number density of dumbbells, $n_\text{d}$. To derive an equation for $n_\text{d}$, we integrate Eq. \eqref{dumbtransp} over the end-to-end vector space. Using integration by parts we obtain:
\begin{equation}\label{dumbconctransp}
\frac{\partial n_\text{d}}{\partial t} + \boldsymbol{v} \cdot \frac{\partial n_\text{d}}{\partial \boldsymbol{x}} = \frac{k_\text{B} T}{2 \xi_\text{d}} \left(\frac{\partial }{\partial \boldsymbol{x}}\right)^2 n_\text{d} + \frac{1}{4 \xi_\text{d}} \frac{\partial }{\partial \boldsymbol{x}} \cdot \left( \frac{d k(c)}{d c} \frac{\partial c}{\partial \boldsymbol{x}} Tr(\boldsymbol{C})\right) \, \, ,
\end{equation}
where in the last term of the right-hand side the scalar product is carried out between the two spatial derivatives. Eq. \eqref{dumbconctransp} yields the evolution of the local number density of dumbbells where the coupling with the solute field is given by the last term on the right-hand side.
We can rewrite the momentum balance using the definition of the conformation tensor as
\begin{equation}\label{mombal2}
 \frac{\partial }{\partial \boldsymbol{x}} \cdot \left[ 2 \eta \, \boldsymbol{D} - P \boldsymbol{I} +   k(c) \boldsymbol{C}-k_\text{B}T n_\text{d} \boldsymbol{I} \right ]=  \boldsymbol{0}\,  \, ,
\end{equation} 
where the term inside the square brackets in the left-hand side can be identified as the divergence of the stress tensor.
The equation that governs the evolution of the solute field remains unchanged by the closure approximation and can be written in terms of $\boldsymbol{C}$ as:
\begin{equation}\label{solbal2}
\frac{\partial c}{\partial t}  = \frac{\partial }{\partial \boldsymbol{x}} \cdot \left( \frac{k_\text{B} T}{\xi} \frac{\partial c}{\partial \boldsymbol{x}} +\frac{c}{2 \xi} \frac{\partial}{\partial \boldsymbol{x}} \left( \frac{d k(c)}{d c} \, Tr(\boldsymbol{C}) \right) - c \boldsymbol{v} \right)\, \, ,
\end{equation} 

In summary, Eqs. \eqref{conftenstransp2} - \eqref{solbal2}, together with the continuity equation Eq. \eqref{cont_eq_solv}, form a closed set of equations for the local number density $n_\text{d}$, the conformation tensor $\boldsymbol{C}$, the solute number density $c$, the solvent velocity $\boldsymbol{v}$ and the pressure $P$, which depend on the spatial coordinate $\boldsymbol{x}$ only. This set of equations can be solved numerically using standard numerical techniques \cite{owens2002computational}. In the Appendix \ref{AppEinst}, we report Eqs. \eqref{conftenstransp2} - \eqref{solbal2} using Einstein index notation.

\subsection{Boundary conditions}
The governing equations given by Eqs. \eqref{conftenstransp2}-\eqref{dumbconctransp} and Eqs. \eqref{mombal1}-\eqref{cont_eq_solv} must be complemented by appropriate boundary conditions. The tangential traction on the volume boundary, $\delta \Omega$, can be specified:
\begin{equation}
\left[\left( 2 \eta \, \boldsymbol{D} - P\, \boldsymbol{I} +  k(c) \boldsymbol{C} -k_\text{B}T n_\text{d} \boldsymbol{I} \right) \cdot \boldsymbol{n}\right] \cdot \left( \boldsymbol{I}- \boldsymbol{n}\boldsymbol{n} \right) = \boldsymbol{\tau}_N \, \, ,
\end{equation}
where $\boldsymbol{\tau}_N$ is the tangential traction imposed on the boundary. This boundary condition is complemented by the condition that the wall is impenetrable to the solvent:
\begin{equation}
\boldsymbol{v} \cdot \boldsymbol{n}=0 \, \, .
\end{equation}
Alternatively, the velocity of the solvent can be specified at the boundary: 
\begin{equation}
\boldsymbol{v} =\boldsymbol{U} \, \, ,
\end{equation}
where $\boldsymbol{U}$ is the velocity at which the boundary is moving. For a stationary boundary $\boldsymbol{U}=\boldsymbol{0}$.

The boundary of the domain is also considered to be impermeable to the solute. It follows that the component of the solute velocity normal to the wall, $\boldsymbol{w} $, vanishes:
\begin{equation}
\boldsymbol{w} \cdot \boldsymbol{n} =0 \, \, ,
\end{equation}
By substituting the solute velocity given by Eq. \eqref{solflux} we obtain:
\begin{equation}\label{boundcondsolv}
-\frac{1}{\xi}\left[k_\text{B} T \frac{\partial c}{\partial \boldsymbol{x}}  +\frac{1}{2} c  \frac{\partial}{\partial \boldsymbol{x}} \left( \frac{d \, k(c)}{d c} Tr(\boldsymbol{C}) \right) \right] \cdot \boldsymbol{n} =0 \, \, .
\end{equation}
%Since the solute transport equation involves fourth-order derivatives we need an additional condition. Here we fix the third order derivative to zero:
%\begin{equation}\label{boundcondsolv1}
%c \frac{\partial}{\partial \boldsymbol{x}} \left(\frac{\partial}{\partial \boldsymbol{x}}\right)^2 c\cdot \boldsymbol{n} =0 \, \, .
%\end{equation}
%The condition above ensures that the distribution of solute is homogeneous at equilibrium and does not vary near the boundaries of the volume.

Finally, the velocity of the dumbbells normal to the wall must vanish:
\begin{equation}
\boldsymbol{w}_d \cdot \boldsymbol{n} =0 \, \, .
\end{equation}
By substituting the dumbbell velocity given by Eq. \eqref{dumbflux1} we obtain:
\begin{equation}
 -\frac{1}{2 \xi_\text{d}}\left( k_\text{B} T \frac{\partial \psi}{\partial \boldsymbol{x}} +\frac{1}{2} \psi  \, \vert\boldsymbol{r}\vert^2  \, \frac{d k(c)}{d c} \frac{\partial c}{\partial \boldsymbol{x}} \right) \cdot \boldsymbol{n} =0 \, \, .
\end{equation}
To obtain the boundary conditions for the local dumbbell number density, $n_\text{d}$,  and for the conformation tensor, $\boldsymbol{C}$, we integrate the equation above over the end-to-end vector space. By integrating over $\boldsymbol{r}$, we obtain
\begin{equation}\label{boundconddumbbconc}
 -\frac{1}{2 \xi_\text{d}}\left( k_\text{B} T \frac{\partial n_\text{d}}{\partial \boldsymbol{x}} +\frac{1}{2} Tr(\boldsymbol{C}) \, \frac{d k(c)}{d c} \frac{\partial c}{\partial \boldsymbol{x}} \right) \cdot \boldsymbol{n} =0 \, \, .
\end{equation}
By multiplying the boundary condition with $\boldsymbol{r}\boldsymbol{r}$, integrating and applying the closure approximation, we obtain:
\begin{equation}\label{boundconddumbbconftens}
 -\frac{1}{2 \xi_\text{d}}\left( k_\text{B} T \frac{\partial }{\partial \boldsymbol{x}} \boldsymbol{C} +\frac{5}{6} \frac{Tr(\boldsymbol{C})}{\Psi} \boldsymbol{C} \, \frac{d k(c)}{d c} \frac{\partial c}{\partial \boldsymbol{x}} \right) \cdot \boldsymbol{n} =0 \, \, .
\end{equation}
In the equation above, the scalar product is carried out between the normal vector and the gradient.

\section{Dimensionless equations}\label{Dimensionless}
To assess the importance of each term appearing in the governing equations in the case of generic flows, we make Eqs.  \eqref{conftenstransp2} - \eqref{solbal2} and Eq. \eqref{cont_eq_solv} dimensionless. We assume that the flows occur over a characteristic length $L$, with a characteristic shear rate $\dot{\gamma}$, so that the characteristic velocity scale is $L \, \dot{\gamma}$. We identify a characteristic dumbbell concentration $n_\text{d,0}$ and a characteristic solute concentration $c_0$, which could be the equilibrium number densities in the absence of flow. Since the spring stiffness of the dumbbells is a function of the local solute number density, we choose a characteristic spring stiffness $k_0$, which we assume to be the stiffness evaluated at $c_0$, $k_0=k(c_0)$. The characteristic stress can be chosen in different ways; we choose to scale it using the solvent viscosity and the characteristic shear rate as $\eta \, \dot{\gamma}$. Finally, the components of the conformation tensor are scaled using $k_\text{B}T \, n_\text{d,0}/k_0$, which is proportional to their value at equilibrium in a homogeneous suspension. 

By plugging these characteristic scales into the governing equations we obtain their dimensionless version, with the dimensionless quantities denoted with a superscript. The dimensionless momentum balance becomes
\begin{equation}\label{mombaldimnless}
\frac{\partial }{\partial \boldsymbol{x}^*} \cdot \left[ 2 \eta^* \, \boldsymbol{D}^* -  \eta^*  \, P^* \boldsymbol{I} + \frac{1}{4 \, De} \left( k^*(c^*) \boldsymbol{C}^* -n_\text{d}^* \boldsymbol{I} \right) \right] =  \boldsymbol{0}\,  \, ,
\end{equation}
where the Deborah number, $De$, compares the characteristic flow timescale with the relaxation time of the dumbbells and it is defined as $De = \dot{\gamma} \xi_\text{d}/4 \, k_0$. In Eq. \eqref{mombaldimnless}, we have introduced the nondimensional viscosity $\eta^*=\eta/\eta_p$ as the ratio of the solvent viscosity and the characteristic polymer viscosity $\eta_p = k_\text{B}T \, n_\text{d,0} \, \xi_\text{d} /k_0$.  The spring stiffness $k^*(c^*)$ and its derivative $\frac{d k^*(c^*)}{d c^*}$ are made dimensionless using the $k_0$ and $c_0$ so that $k^*(1)=1$. %, and the dimensionless nonlocal interaction between the solute molecules, $\Lambda^*$, as $\Lambda^*= \Lambda \, c_0/k_\text{B}T \, L^2 $. This dimensionless number is typically very small because the scale over which the solute molecules interact is typically much smaller than the characteristic lengthscale of the flow. It has a limited impact to flows where the characteristic length over which the solute distribution changes is macroscopic. Nevertheless, this term plays an important role in the case of equilibrium or nonequilibrium phase separation because it stabilizes the boundaries between the two phases.
The dimensionless continuity equation is unchanged
\begin{equation}\label{cont_eq_solv_dimless}
\frac{\partial }{\partial \boldsymbol{x}^*} \cdot \boldsymbol{v}^*=0 \, \, .
\end{equation}
The dimensionless equation of transport of solute reads:
\begin{equation}\label{soltranspdimless}
\frac{\partial c^*}{\partial t^*} +  \boldsymbol{v}^* \cdot \frac{\partial}{\partial \boldsymbol{x}^*} c^* = \frac{\partial }{\partial \boldsymbol{x}^*} \cdot \left( \frac{1}{Pe} \frac{\partial c^*}{\partial \boldsymbol{x}^*} +\frac{n_0^*}{2 \,  Pe} c^*\frac{\partial}{\partial \boldsymbol{x}^*} \left( \frac{d k^*(c^*)}{d c^*} \, Tr\left(\boldsymbol{C}^*\right) \right) \right)\, \, ,
\end{equation} 
where we have introduced the solute P\'{e}clet number as $Pe= \dot{\gamma} \, L^2 \, \xi/k_\text{B}T$ and  the ratio of the dumbbell and solute characteristic concentrations $n_0^*=n_\text{d,0}/c_0$. The P\'{e}clet number compares the importance of solute transport due to fluid flow to that due to diffusion. 
The evolution of the dimensionless dumbbell number density is given by
\begin{equation}\label{dumbconctranspdimless}
\frac{\partial n_\text{d}^*}{\partial t^*} + \boldsymbol{v}^* \cdot \frac{\partial n_\text{d}^*}{\partial \boldsymbol{x}^*} = \frac{1}{2 \, Pe_d} \left(\frac{\partial }{\partial \boldsymbol{x}^*}\right)^2 n_\text{d}^* + \frac{1}{4 \, Pe_d} \frac{\partial }{\partial \boldsymbol{x}^*} \cdot \left( \frac{d k^*(c^*)}{d c^*} \frac{\partial c^*}{\partial \boldsymbol{x}^*} Tr(\boldsymbol{C}^*)\right) \, \, ,
\end{equation}
where we defined the dumbbell P\'{e}clet number as $Pe_d= \dot{\gamma} \, L^2 \, \xi_\text{d}/k_\text{B}T$.
Finally, employing the closure approximation described in section \ref{closureapproximation}, the equation governing the evolution of the dimensionless conformation tensor reads:
\begin{equation}\label{conftenstranspdimless}
De \,\overset{\nabla}{\boldsymbol{C}^*}= -\left( k^*(c^*) \boldsymbol{C}^*-n_\text{d}^* \boldsymbol{I}\right) + \frac{De}{2 \, Pe_d} \left(\frac{\partial }{\partial \boldsymbol{x}^*}\right)^2 \boldsymbol{C}^* + \frac{5 \, De}{12 Pe_d} \frac{\partial }{\partial \boldsymbol{x}^*} \cdot \left( \frac{d k^*(c^*)}{d c^*} \frac{\partial c^*}{\partial \boldsymbol{x}^*} \frac{Tr(\boldsymbol{C}^*)}{\Psi^*} \boldsymbol{C}^*\right) \, \, .
\end{equation}
Let us now assess the validity of the closure approximation that we introduced in the previous section. Eq. \eqref{conftenstranspdimless} shows that the spatial flux of the conformation tensor is proportional to $De/Pe_d= k_\text{B}T/4 k_0 L^2$. This ratio is independent of the shear rate and depends on the dumbbell relaxation time, on its friction coefficient and on the characteristic length of the problem only. It can be interpreted as the ratio between the time required to diffuse over a length $L$ and the relaxation time of the dumbbell. Typically, the relaxation time of a polymer is in the order of $1 \rm{s}$ and its diffusion coefficient is in the order $1 \rm{\mu m}^2 \rm{s}^{-1}$. Therefore, unless the characteristic lengthscale of the problem is smaller than about $1 \rm{\mu m}$, dumbbells relax faster than they diffuse, which results in $De/Pe_d \ll 1$. As a consequence, we expect that in many cases the closure approximation will have a limited impact on the fluid flow.

\section{Shear-induced phase separation}\label{SIPS}
\begin{figure}[h!]
\centering
\includegraphics[width=0.75\textwidth]{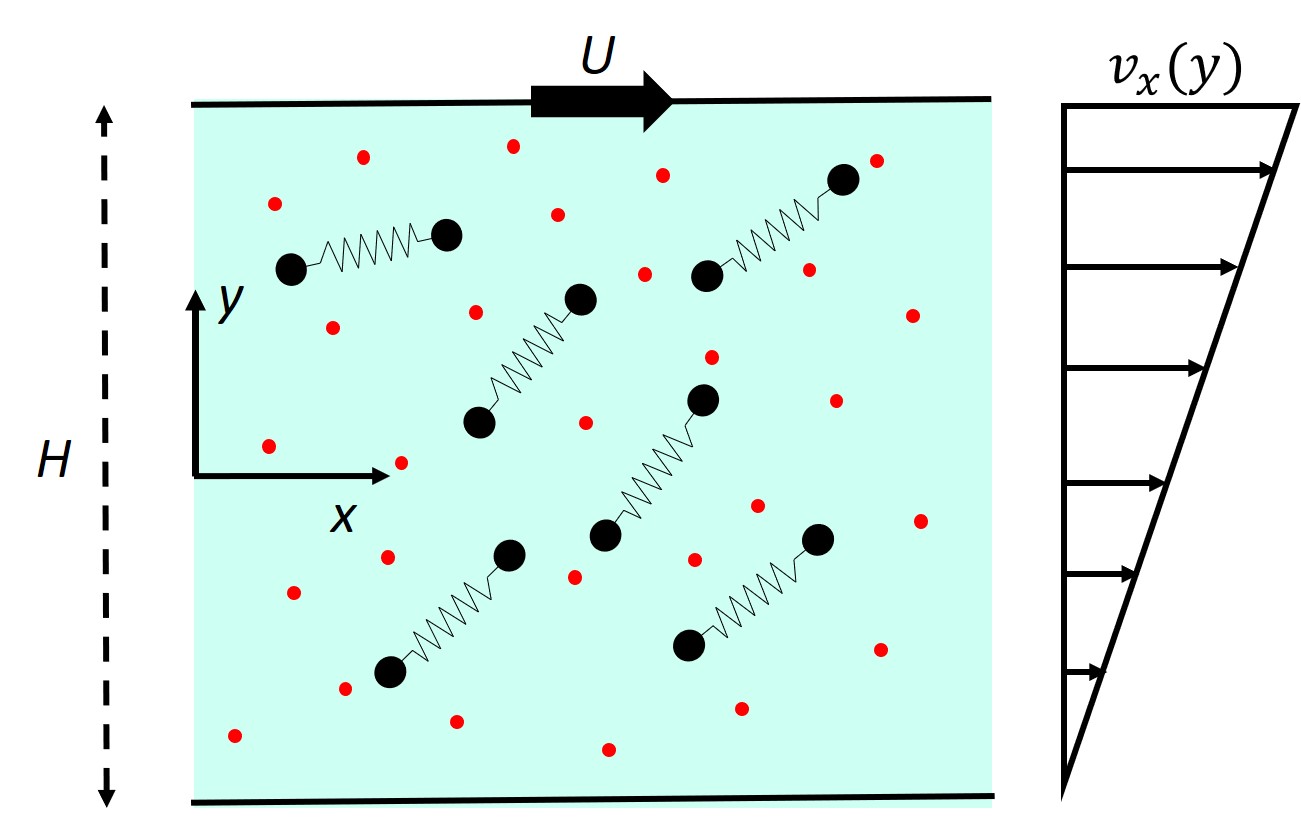}
\caption{Schematics of the homogeneous shear state. The dumbbells and the solute are homogeneously distributed along $y$ and the shear rate is constant.}
\label{fig4}
\end{figure}

%Experiments considering a suspension of linear PNIPAM polymers in a ionic solution showed that applying an external shear lead to a shear-induced phase separation whereby an increase of the turbidity of the suspension is observed [ref]. 
Motivated by the experimental evidence of shear-induced phase separation \cite{Larson_2005}, we study the stability of a homogeneously sheared suspension of solute and dumbbells to small perturbations. We focus on homogeneous suspensions of solute, dumbbells and solvent that are thermodynamically stable and we investigate the possibility of a phase separation that is promoted by the shear flow. We address this question using linear stability analysis and a simplified theory. As shown schematically in Figure \ref{fig4}, we consider perturbations around a homogeneous suspension of solute and dumbbells sheared between two parallel walls. We identify the characteristic dumbbell and solute number densities as those in the homogeneous state $n_\text{d,0}$ and $c_0$ in the homogeneous state. The characteristic dumbbell spring stiffness is given by $k(c_0)$. The walls are separated by a distance $H$ and the top wall is moved along the $x$ axis at a speed $U$. In this case, we identify $H$ as characteristic length, and as characteristic shear rate, $\dot{\gamma}=U/H$. 
We consider small perturbations to the homogeneous shear state and we assume that the perturbed variables only change along the $y$ axis. We denote the perturbations with a prime:
\begin{equation}\label{perturbshear}
c^*=1+c^{*'} \, , \, \, \, n_\text{d}^*=1+n_\text{d}^{*'} \,, \, \, \, v^*_x=(y^*+1/2) +v^{*'}_x \,,  \, \, \,  \boldsymbol{C}^*=\boldsymbol{C}_0^*+\boldsymbol{C}^{*'} \, \, .
\end{equation}
In the homogeneous base state the shear rate, the number densities of solute and dumbbells and the conformation tensor are constant along in the $y$ direction. Under these conditions, the governing equations reduce to the Oldroyd-B model, which predicts the components of the conformation tensor, $\boldsymbol{C}_0^*$, to be given by $C^*_{xx,0}=1+2 \, De^2$, $C^*_{yy,0}=C^*_{zz,0}=1$ and $C^*_{xy,0}=C^*_{yx,0}= De$.
The linear stability analysis proceeds in the standard way. First, we substitute the perturbation in the governing equations and in the boundary conditions retaining the linear terms only. The resulting set of equations is a system of linear partial differential equations, which can be solved using eigenfunctions appropriate for the geometry considered here. 

By plugging the perturbation in the dimensionless governing equations, given by Eqs \eqref{mombaldimnless}-\eqref{conftenstranspdimless}, neglecting the nonlinear terms and Taylor expanding the function $k^*(c^*)$ around $c^*=1$, we obtain the equations governing the evolution of the perturbations.
The evolution of the solute perturbation reads
\begin{equation}\label{perturbshearconcsol}
\frac{\partial c^{*'}}{\partial t^*} = \frac{1}{Pe}\frac{\partial }{\partial y^*} \bigg[ \frac{\partial c^{*'}}{\partial y^*} +\frac{n_0^*}{2 } \frac{\partial}{\partial y^*} \bigg( \frac{d k^*(c^*)}{d c^*}\Bigr|_{c^*=1} \, Tr\left(\boldsymbol{C}^{*'}\right) + Tr(\boldsymbol{C}_0^*) \, \frac{d^2 k^*(c^*)}{d c^{*2}}\Bigr|_{c^*=1} \, c^{*'} \bigg) \bigg]\, \, ,
\end{equation}
where we used the fact that the perturbations to the solute concentration only vary along the $y$ axis. The evolution of the perturbations to the dumbbell number density, $n_\text{d}^{*'}$, is given by
\begin{equation}
\frac{\partial n_\text{d}^{*'}}{\partial t^*} = \frac{1}{2 \, Pe_d}\frac{\partial }{\partial y^*} \bigg[ \frac{\partial n_\text{d}^{*'}}{\partial y^*} + \frac{1}{2 } Tr(\boldsymbol{C}_0^*) \, \frac{d k^*(c^*)}{d c^*}\Bigr|_{c^*=1} \,   \frac{\partial}{\partial y^*} c^{*'} \bigg]\, \, .
\end{equation}
Since the velocity is directed along the $x$ axis and the perturbations vary along the $y$ axis only, there is no contribution to the transport of solute or of dumbbells due to advection.
By substituting the perturbations into Eq. \eqref{conftenstranspdimless}, we obtain an equation for the evolution of perturbation to the conformation tensor: 
\begin{multline}\label{perturbconformationtensor}
De \,\left[ \frac{\partial }{\partial t^*}\boldsymbol{C}^{*'} -\frac{\partial }{\partial y^*}v^{*'}_x \left( \hat{\boldsymbol{x}}\hat{\boldsymbol{y}} \cdot \boldsymbol{C}_0^{*} + \boldsymbol{C}_0^{*} \cdot \hat{\boldsymbol{y}}\hat{\boldsymbol{x}}  \right) - \left( \hat{\boldsymbol{x}}\hat{\boldsymbol{y}} \cdot \boldsymbol{C}^{*'} + \boldsymbol{C}^{*'} \cdot \hat{\boldsymbol{y}}\hat{\boldsymbol{x}}  \right)  \right]= \\ =
-\left(\boldsymbol{C}^{*'}+\boldsymbol{C}_0^* \frac{d k^*(c^*)}{d c^*}\Bigr|_{c^*=1} \, c^{*'}-n_\text{d}^{*'} \boldsymbol{I}\right) + \frac{De}{2 \, Pe_d} \frac{\partial^2 }{\partial y^{*2}} \boldsymbol{C}^{*'} + \frac{5 \, De}{12 Pe_d} \frac{d k^*(c^*)}{d c^*}\Bigr|_{c^*=1} \frac{\partial }{\partial y^*} \left( \frac{\partial c^{*'}}{\partial y^*} Tr(\boldsymbol{C}_0^*) \boldsymbol{C}_0^*\right) \, \, .
\end{multline}
where $\hat{\boldsymbol{x}}$ and $\hat{\boldsymbol{y}}$ are the unit vectors along the $x$ and $y$ axis, respectively. Since the conformation tensor is symmetric, there are four independent components that are nonzero: the three diagonal components and the $xy$ component. 
Substitution of the perturbed quantities in the momentum balance yields for the $x$ component
\begin{equation}\label{perturbshearxmombal}
\frac{\partial }{\partial y^*}  \left( \eta^* \, \frac{\partial }{\partial y^*} v^{*'}_x + \frac{1}{4 \, De} C^{*'}_{xy} + \frac{1}{4} \frac{d k^*(c^*)}{d c^*}\Bigr|_{c^*=1} \, c^{*'} \right) = 0  \,  \, ,
\end{equation}
in which only the $xy$ component of the conformation tensor appears. The $y$ component of the momentum balance simply states that the pressure balances variations of $C^{*'}_{yy}$ and of the nonlocal solute interactions along the $y$ axis.
Finally, since the perturbed velocity field $v^{*'}_x$ varies along the $y$ axis only, the continuity equation is identically satisfied.

We assume that the walls, positioned at $y^*=\pm 1/2$, are impermeable to the solute, to the dumbbells and to the solvent. The tangential velocity of the solvent on the top wall, $y^*=1/2$, is fixed and it is given by $\boldsymbol{U}^*=\hat{\boldsymbol{x}}$. The boundary conditions for the perturbations are obtained by making Eq. \eqref{boundcondsolv} and Eqs. \eqref{boundconddumbbconc}-\eqref{boundconddumbbconftens} dimensionless and plugging in the perturbations given by Eqs. \eqref{perturbshear}. The boundary condition at $y^*=\pm 1/2$ for the perturbation to the solute, $c^{*'}$ reads
%\begin{equation}\label{concpertshearbc1}
%\frac{\partial^3}{\partial y^{*3}} c^{*'}=0 \, \, ,
%\end{equation}
%and
\begin{equation}\label{concpertshearbc2}
\frac{\partial c^{*'}}{\partial y^*} +\frac{1}{2}  \frac{\partial}{\partial y^*} \left( \frac{d k^*(c^*)}{d c^*}\Bigr|_{c^*=1} Tr(\boldsymbol{C}^{*'})+\frac{d^2 k^*(c^*)}{d c^{*2}}\Bigr|_{c^*=1}  \, Tr(\boldsymbol{C}_0^{*})\, c^{*'} \right)  =0 \, \, .
\end{equation}
Since the tangential velocity is fixed at the top and bottom walls, the perturbations must vanish
\begin{equation}
v_x^{*'}=0 \, \, .
\end{equation}
The perturbation to the dumbbell number density satisfies 
\begin{equation}\label{dumbpertshearbc1}
\frac{\partial n_\text{d}^{*'}}{\partial y^*} +\frac{1}{2} Tr(\boldsymbol{C}_0^{*}) \, \frac{d k^*(c^*)}{d c^*}\Bigr|_{c^*=1} \frac{\partial c^*}{\partial y^*}  =0 \, \, ,
\end{equation}
and the perturbation to the conformation tensor satisfies
\begin{equation}\label{boundconddumbbconftens1}
 \frac{\partial }{\partial y^*} \boldsymbol{C}^{*'} +\frac{5}{6} Tr(\boldsymbol{C}_0^{*}) \boldsymbol{C}_0^{*} \, \frac{d k^*(c^*)}{d c^*}\Bigr|_{c^*=1} \,  \frac{\partial c^{*'}}{\partial y^*}  =0 \, \, ,
\end{equation}
at $y^*=\pm 1/2$.

The linear stability analysis now proceeds in the standard way \cite{leal2007advanced}. We assume that the spatial and temporal dependence of the perturbed quantities can be separated. The spatial variations of the perturbations can be expanded in Fourier series and the temporal dependence is simply given by an exponential.
To satisfy the no flux boundary conditions, given by Eqs \eqref{concpertshearbc2}-\eqref{boundconddumbbconftens1} the perturbation of the concentration, dumbbell density and conformation tensor must be given by cosine series
\begin{equation}\label{fourierconc}
c^{*'}= \sum_{n=1}^{\infty} c^{*'}_n \cos{(2 \pi n y^*)} \, \exp{(\lambda_n \, t^*)} \, \, \, ,
\end{equation}
\begin{equation}
n_\text{d}^{*'}= \sum_{n=1}^{\infty} n^{*'}_{\text{d},n} \cos{(2 \pi n y^*)} \, \exp{(\lambda_n \, t^*)} \, \, \, ,
\end{equation}
and
\begin{equation}
\boldsymbol{C}^{*'}= \sum_{n=1}^{\infty} \boldsymbol{C}^{*'}_n \cos{(2 \pi n y^*)} \, \exp{(\lambda_n \, t^*)} \, \, \, ,
\end{equation}
where $c^{*'}_n$, $n^{*'}_{\text{d},n}$ and $\boldsymbol{C}^{*'}_n$ is the initial value of the perturbation with mode $n$ and $\lambda_n $ is its dimensionless growth rate. 
The perturbation to the velocity, which satisfies no-slip boundary conditions on the upper and bottom walls, is given by a sine series
\begin{equation}\label{fouriervel}
v^{*'}_x= \sum_{n=1}^{\infty} v^{*'}_{x,n} \sin{(2 \pi n y^*)} \, \exp{(\lambda_n \, t^*)} \, \, \, ,
\end{equation}
where $v^{*'}_{x,n}$ is the initial value of a velocity perturbation of mode $n$. It is straightforward to verify that the ansatz, given by Eqs \eqref{fourierconc}-\eqref{fouriervel}, satisfies the boundary conditions. In the Appendix \ref{appB} it is shown that, by inserting Eqs \eqref{fourierconc}-\eqref{fouriervel} into Eqs \eqref{perturbshearconcsol}-\eqref{perturbshearxmombal}, we obtain a set of eigenvalue problems for each growth rate $\lambda_n$. Since there are six independent variables, for each Fourier mode we find six eigenvalues. The homogeneous shear state is unstable if any of the eigenvalues has a positive real part. As shown in the Appendix  \ref{appB}, the eigenvalue problem for each mode $n$ can be written in a compact form as:
\begin{equation}\label{eig_probl_comp}
\lambda_n \, X - E \cdot X =0 \, \, ,
\end{equation}
with the vector $X$ being the vector of the initial values of the perturbations:
\begin{equation}
X = \begin{bmatrix} c^{*'}_n\\ n^{*'}_{\text{d},n} \\ C_{xx,n}^{*'} \\ C_{yy,n}^{*'} \\ C_{zz,n}^{*'} \\ C_{xy,n}^{*'} \end{bmatrix} \, \, ,
\end{equation}
and $E$ being a six by six matrix. 
The values of $\lambda_n$ that satisfy Eq. \eqref{eig_probl_comp} are given by the eigenvalues of the matrix $E$. In our case, there are six eigenvalues for each mode $n$, one for each field that is being perturbed. To investigate the stability of the homogeneous state for a given set of parameters, we have to solve the eigenvalue problem numerically for each mode $n$.  We verified that, by solving the eigenvalue problem given by Eq. \eqref{eig_probl_comp} for very small values of $De$, we correctly recover the results obtained in Section \ref{equiPS} for a solution at equilibrium. This is expected because for $De \ll 1$ the solution is essentially at equilibrium and we expect to recover the behavior studied in Section \ref{equiPS}.

Nevertheless, analytical progress can be made by realizing that in typical situations some dimensionless number is very large or very small. In typical experimental conditions the gap between the walls is $H\approx 1 \rm{mm}$, the shear rate of the order of $\dot{\gamma} \approx 1 \rm{s}^{-1}$ and the dumbbell diffusion coefficient $k_\text{B}T/\xi_\text{d} \approx 1 \rm{\mu {m}^2} \, \rm{s}^{-1}$, which makes the dumbbell P\'{e}clet number, $Pe_d\approx 10^6$, much larger than one. The diffusion coefficient of the solute molecules, $k_\text{B}T/\xi \approx 10^3 \rm{\mu {m}^2} \, \rm{s}^{-1}$, is much larger than that of the dumbbells, therefore it is $Pe/Pe_d \ll 1$. As a result, there is a separation of timescales between the evolution of the perturbations of the solute number density and that of the dumbbells. It follows that the number density of dumbbells remains essentially constant over the evolution of the perturbations. Similarly, there is also a separation of timescales between the relaxation time of the dumbbells and the time required by the solute and by the dumbbells to migrate over the length of the gap: $De/Pe_d \ll 1$ and $De/Pe \ll 1$. It follows that the conformation of the dumbbells relaxes quasi-statically to the value determined by the local shear rate and by the value of the spring stiffness. Finally, we consider as a simplification that the relative viscosity is very small $\eta^* \ll 1$, which is typically the case of polymer solutions.
Under these assumptions, we consider that the dumbbell concentration remains homogeneous $n_\text{d}^{*'}=0$, that the time derivatives of the conformation tensor are zero and that $De/Pe_d=0$. By doing so (see Appendix  \ref{appC}), we find that there exists a critical Deborah number above which the homogeneous shear is unstable
\begin{equation}\label{decrit}
De_\text{crit} = \left( \frac{1 }{n_0^* \left[ \left( \frac{d k^*(c^*)}{d c^*}\Bigr|_{c^*=1} \right)^2 - \frac{d^2 k^*(c^*)}{d c^{*2}}\Bigr|_{c^*=1} \right]} - \frac{3}{2}\right)^{1/2} \, \, .
\end{equation}
There are some similarities with the equilibrium phase separation studied in Section \ref{equiPS}. The critical Deborah number depends on the dimensionless spring stiffness slope squared, $(d k^*(c^*)/d c^*\Bigr|_{c^*=1})^2$, which means that it is not relevant whether the solute makes the dumbbells stiffer or softer only how sensitive is the stiffness with respect to changes of solute concentration. Conversely, the sign of the second derivative, $d^2 k^*(c^*)/d c^{*2}\Bigr|_{c^*=1}$, is relevant: a positive curvature of the function $k^*(c^*)$ at $c^*=1$ hinders the instability while a negative sloe promotes the instability. Interestingly, similarly to the case of the equilibrium phase separation, the instability threshold does not depend on transport coefficients.
Some general conclusions can be drawn for mixtures of dumbbells and solute in which the spring constant is a power law of the local solute number density, $k(c) \propto \, c^N$. In this case, the dimensionless derivative, $(d k^*(c^*)/d c^*)^2=N^2$ and the dimensionless curvature, $d^2 k^*(c^*)/d c^{*2}=N(N-1)$, are constant and depend only on the power-law exponent $N$. It follows that the term inside the square bracket of the denominator of Eq. \eqref{decrit} is simply equal to $N$ and it does not depend on the prefactor of the power law. These insights suggest that the shear-induced instability cannot occur in the case of the spring constant being a decreasing power law of the solute number density, $N<0$. One system of chemically-responsive polymers that displays $k(c) \propto \, c^N$ with $N=1$ is a mixture of polyelectrolytes and salt. These systems often display a shear-banding instability and the mechanism described here could play a role in the transition from the homogeneous to the shear-banded regimes. 

\begin{figure}[h!]
\centering
\includegraphics[width=1.0\textwidth]{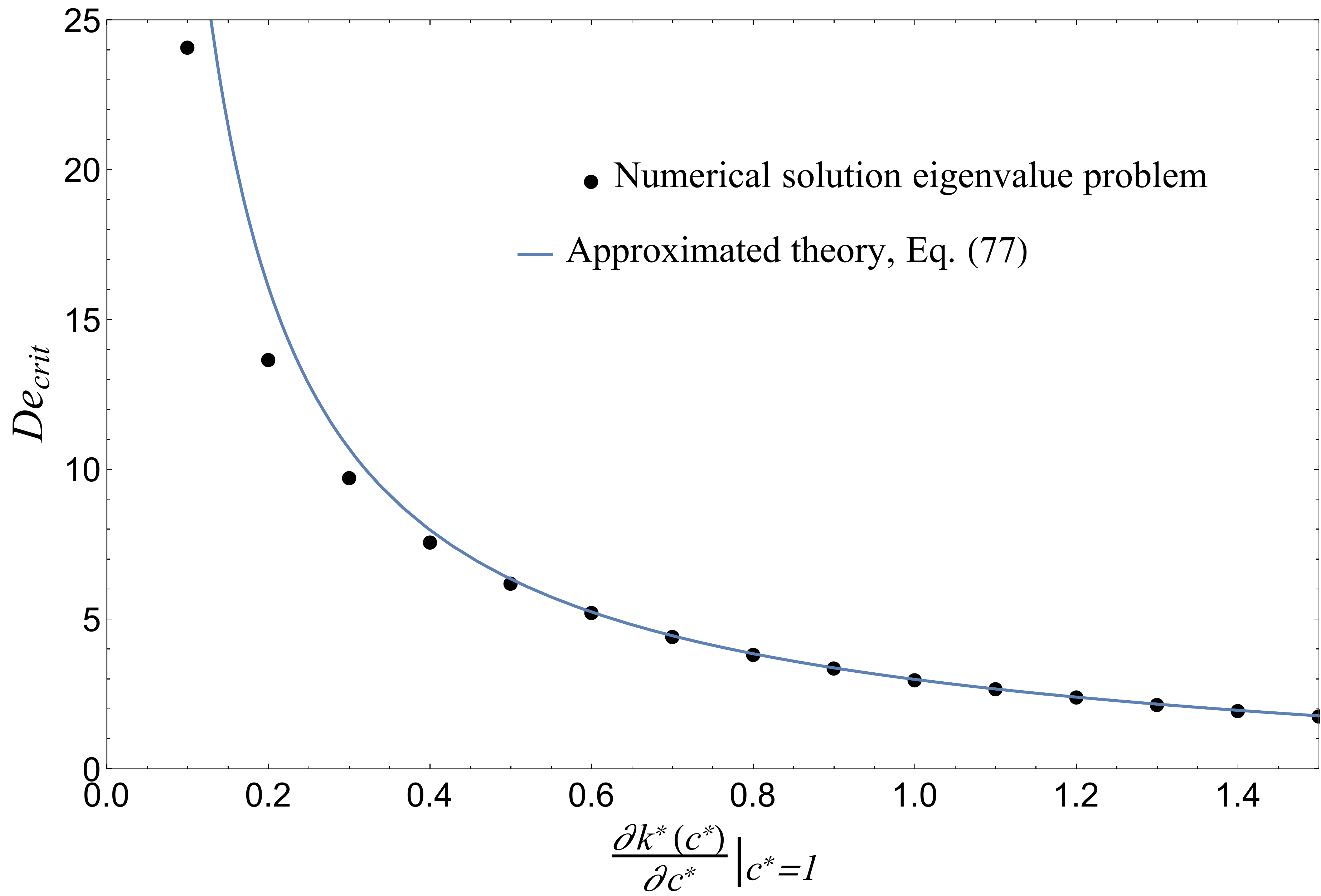}
\caption{The critical Deborah number obtained from the numerical solution of the eigenvalue problem (solid circles) and from the simplified theory given by Eq. \eqref{decrit}. The other dimensionless parameters are $Pe_d=10^6$, $Pe=10^3$, $\eta^*=10^{-3}$, $n_0^*=0.1$ and for $d^2 k^*(c^*)/d c^{*2}\Bigr|_{c^*=1}=0$. }
\label{fig6}
\end{figure} 

In Figure \ref{fig6}, we compare the critical Deborah number predicted by the simplified expression, given by Eq. \eqref{decrit}, and that obtained from the numerical solution of the eigenvalue problem. To find the critical Deborah number numerically, we proceed iteratively. We fix the dimensionless parameters and $De=0.01$ and look for the eigenvalues for the Fourier modes from $n=1$ up to a large threshold value fixed to $n=10000$. If all the eigenvalues have a negative real part, we increase $De$ by a $0.01$ and we repeat the process until we find the critical Deborah number $De=De_\text{crit}$ for which at least one eigenvalue has a positive real part. In Figure \ref{fig6}, we show the results for $Pe_d=10^6$, $Pe=10^3$, $\eta^*=10^{-3}$, $n^*=0.1$ and for $d^2 k^*(c^*)/d c^{*2}\Bigr|_{c^*=1}=0$, for which the approximated theory is expected to be valid. The results show that the value of $De_\text{crit}$ given by Eq. \eqref{decrit} yields a very good approximation of that obtained from the numerical solution. The small deviation at small values of $d k^*(c^*)/d c^{*}\Bigr|_{c^*=1}$ is due to the large critical Deborah number. The spatial flux of dumbbells scale as $De^2$, which makes it very large when $De_\text{crit}\gg 1$. At such large Deborah numbers the perturbations of the dumbbell number density, which are neglected in the simple analytical theory, become important. 

\begin{figure}[h!]
\centering
\includegraphics[width=1.0\textwidth]{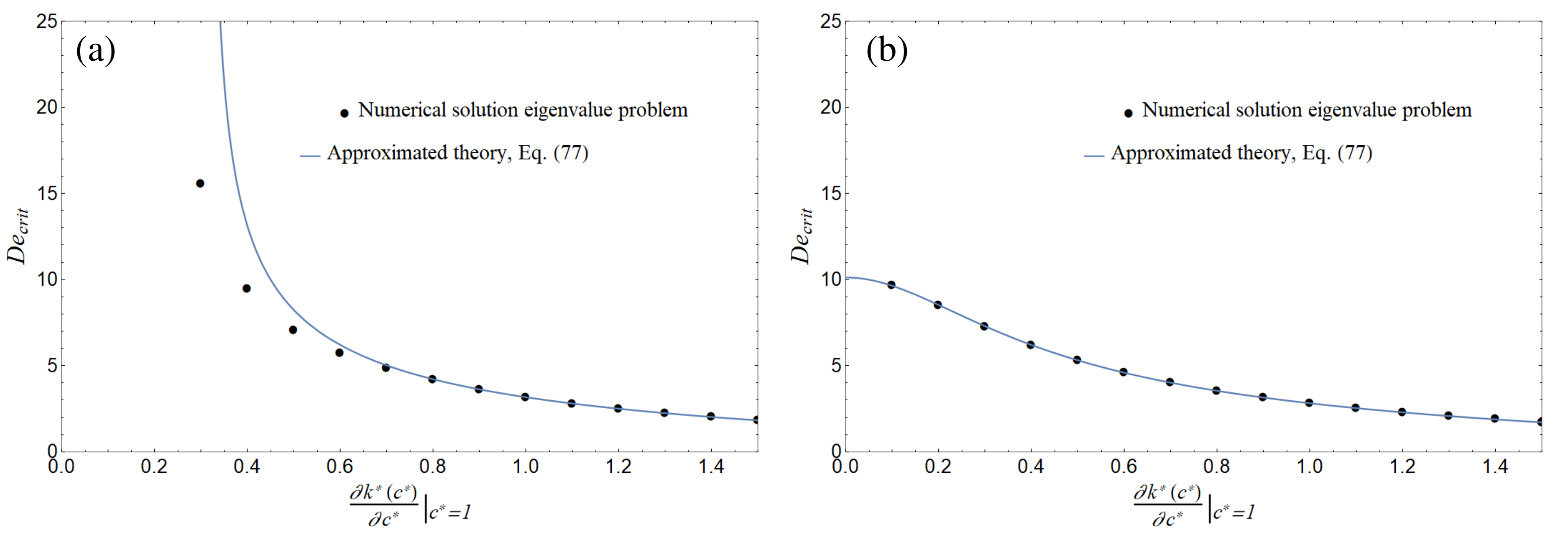}
\caption{The critical Deborah number obtained from the numerical solution of the eigenvalue problem (solid circles) and from the simplified theory given by Eq. \eqref{decrit}. Panel (a) shows the case of $d^2 k^*(c^*)/d c^{*2}\Bigr|_{c^*=1}= 0.1$, while Panel (b) shows the case of $d^2 k^*(c^*)/d c^{*2}\Bigr|_{c^*=1}= -0.1$ The other dimensionless parameters are $Pe_d=10^6$, $Pe=10^3$, $\eta^*=10^{-3}$, $n_0^*=0.1$ and for $d^2 k^*(c^*)/d c^{*2}\Bigr|_{c^*=1}=0$. }
\label{fig7}
\end{figure}

\begin{figure}[h!]
\centering
\includegraphics[width=1.0\textwidth]{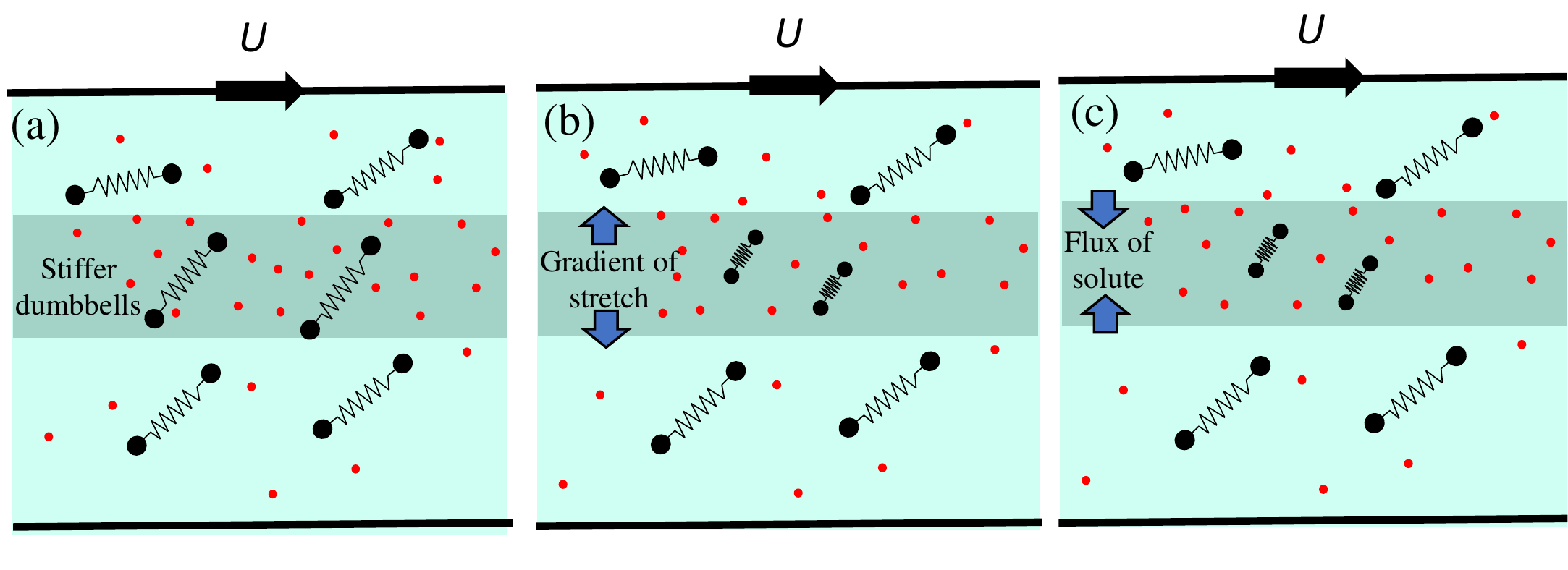}
\caption{Schematics of the shear-induced phase separation mechanism for the case of dumbbells whose spring stiffness increases with an increased concentration of solute. (a) A perturbation of the solute concentration leads to a local accumulation of solute in the shaded region. In the case of $d k^*(c^*)/d c^*\Bigr|_{c^*=1} >0$, this results in a local increase of the dumbbell spring stiffness. (b) As a result of the local stiffness increase in the shaded region, the dumbbells contract, thus creating a gradient of stretch along the $y$ coordinate. (c) The spatial variation of the dumbbell stretch drives a solute flux towards the region with a larger density of solute, leading to a positive feedback loop.}
\label{fig5}
\end{figure}

In Figure \ref{fig7}, we show the comparison between the theory and the simulations for the case of $d^2 k^*(c^*)/d c^{*2}\Bigr|_{c^*=1}= \pm 0.1$. A positive curvature of the function $k^*(c^*) $ at $c^*=1$ hinders the instability while a negative curvature promotes it. For a negative curvature, the instability can occur even if the slope is zero. Since the perturbations of the dumbbell density are ignored in the derivation of Eq. \eqref{decrit}, the agreement between the numerical solution and the simplified theory suggests that the instability is driven by a positive feedback mechanism between the solute migration and the stretch of the polymers. The dumbbell migration is not involved in the instability mechanism and it occurs over longer timescales, therefore being relevant for the final steady-state reached by the system only. The positive feedback between the solute field and the dumbbell conformation is depicted schematically in Figure \ref{fig5} and takes place as follows. A perturbation of the solute density leads to a local accumulation in certain regions and local depletion in others (Figure \ref{fig5}(c)). The solute perturbation changes the stiffness of the dumbbells, which, in turn, relax quickly to the new conformation. In the case $d k^*(c^*)/d c^*\Bigr|_{c^*=1}>0$, the dumbbells contract where the solute concentration has increased and they stretch where the solute concentration has decreased (Figure \ref{fig5}(b)). This results in a gradient of dumbbell stretch along the gap, which drives solute from areas where the dumbbells are more stretched to areas where they are contracted (Figure \ref{fig5}(c)). The solute flux reinforces the initial perturbation, ultimately leading to an instability. It is straightforward to show that the growth of the initial perturbations is driven by a decrease in the total elastic energy. In dimensionless form, the elastic energy is written as 
\begin{equation}
\int_{y^*=-1/2}^{y^*=1/2} k^*(c^*) \, Tr(\boldsymbol{C}^*) dy^* \, \, .
\end{equation}
By expanding the elastic energy up to second order in the perturbations, we obtain
\begin{equation}\label{elasticenergyperturb}
\int_{y^*=-1/2}^{y^*=1/2}  \frac{d k^*(c^*)}{d c^*}\Bigr|_{c^*=1} c^{*'} \, Tr(\boldsymbol{C}^{*'} )dy^* \, \, .
\end{equation}
Since the dumbbells relax very quickly to equilibrium and their concentration evolves slowly, by using Eq. \eqref{perturbconformationtensor} we find that $Tr(\boldsymbol{C}^{*'}) \approx  -Tr(\boldsymbol{C}_0^{*}) \frac{d k^*(c^*)}{d c^*}\Bigr|_{c^*=1} c^{*'}$. By plugging this into the expansion of the elastic energy, given by Eq. \eqref{elasticenergyperturb}, we obtain
\begin{equation}
\int_{y^*=-1/2}^{y^*=1/2}  -Tr(\boldsymbol{C}_0^{*'} ) \left(\frac{d k^*(c^*)}{d c^*}\Bigr|_{c^*=1} \right)^2  c^{*'2} \, dy^* \, \, .
\end{equation}
Since the trace of the conformation tensor is a positive quantity representing the extension of the dumbbells and the solute perturbations, $c^{*'}$, are given by a Fourier series, it is straightforward to show that the integral is always negative. It follows that any perturbation to the solute concentration reduces the total elastic energy of the system.
%\marino{[I did not have too much time to think about it, but more stretched polymers in the margins in the figure are also softer and hence it is not obvious if they have larger elastic energy or not, which is what they try to minimize. So, I am not fully convinced by this explanation. Am I missing something?]} 
The very same mechanism described above takes place in the case $d k^*(c^*)/d c^*\Bigr|_{c^*=1}<0$.  

Previous works showed that instabilities in shear flow can be produced mainly through two mechanisms: (i) a nonmonotonic stress-strain rate relation \cite{Fielding_2014,Germann_2019} and (ii) polymer migration driven by inhomogeneous elastic stress \cite{Helfand_1989,Doi_1992,Cromer_2013,Cromer_2014,Peterson_2016,Peterson_2020,Burroughs_2021}. In contrast, our work identifies a new mechanism for shear instability that couples the accumulation of polymers and solute molecules. This mechanism could be the dominant one if the conformation of a polymer is highly sensitive to the concentration of solute molecules.

\section{Conclusions}
In this paper, we present a theory for the flow of a suspension of chemically-responsive polymers that can change conformation depending on the local concentration of a solute as is the case, for instance, of polyelectrolytes and salt. To capture the solute-dependent conformation changes while keeping the model as simple as possible, we assume that the polymers can be idealized as dumbbells linked by a linear spring whose stiffness depends on the local number density of a solute. The choice of a linear dumbbell model is expected to be a good description of dilute and semi-dilute polymer suspensions in the case of polymer chains that are not fully stretched. The theory is valid for spatially inhomogeneous dumbbells and solute concentration. % There are two mechanisms that impact the dumbbell extension: (i) local change of solute concentration that change the spring stiffness, and (ii) the dumbbell stretch generated by straining flows. 

%NOTE THAT THIS IS DIFFERENT FROM CONSIDERING THE CASE OF A POOR/GOOD SOLVENT. BECAUSE...SOLVENT IS INCOMPRESSIBLE. SOLUTE CAN CHANGE CONCENTRATION AND BE HIGHLY INHOMOGENEOUS.

To derive the equation governing the flow of a mixture of solute and dumbbells, we employ the Onsager's variational formalism. By doing so, we obtain the balance of solute, dumbbells and momentum density directly from the rate of change and the rate of dissipation of the free energy. The governing equations unveil novel couplings between the solute and the dumbbell. We find that spatial gradients of dumbbell extension, which are ubiquitous in flows with curved streamlines, drive fluxes of solute. Likewise, gradients of solute concentration lead to spatial fluxes of dumbbells and vice versa. This mechanism of dumbbell migration is fundamentally different from previously proposed mechanisms of polymer migration due to curved streamlines \cite{Aubert_1980,Aubert_1980_1} or spatial variations of elastic stresses  \cite{Helfand_1989,Doi_1992,Cromer_2013,Tsouka_2014}.
As a consequence of these couplings, in the case of a generic flow, the densities of solute and dumbbells are spatially-inhomogeneous and so are the rheological properties of the mixture.

The distribution of dumbbells is determined by a partial differential equation defined in a six-dimensional space consisting of their spatial coordinate in a fixed frame and of the end-to-end vector coordinate. This feature makes it difficult to solve the set of governing equations numerically, especially under generic flow conditions. To circumvent this problem, we propose a closure approximation that allows computing directly the second moment of the dumbbell configuration, which is then used in the remaining equations. This approximation eliminates the requirement to compute the entire distribution of dumbbells and leads to a set of equations that only depend on the spatial coordinate, thus being more amenable to numerical discretization. Using dimensional analysis, we show that the closure approximation introduces only a minimal error for flows where the relaxation time of the dumbbells is much smaller than the time required to diffuse over the characteristic length.

Using a linear stability analysis of a homogeneous mixture of solute and dumbbells around the equilibrium, we find that the solution can undergo a thermodynamic phase separation. The homogeneous solution is unstable in the case of rapid variations of the spring stiffness as a function of the solute concentration. When the mixture phase separates, the dumbbells aggregate in regions of small(high) solute concentration if increasing its concentration leads to stiffer(softer) springs. %This phase transition is driven by a total increase of entropy associated with the increased conformational microstates available to the softer dumbbells, which overcomes the reduction of entropy due to the spatial organization.
Similar phase transitions have been observed in stimuli-responsive polymer suspensions that undergo a coil to globule conformation transition.

We investigate the stability of a mixture of solute and dumbbells that is sheared between two parallel walls. We find that the homogeneous state undergoes spontaneous demixing in the case of sufficiently large shear rates. Using a linear stability analysis, we derive an approximated expression of the critical Deborah number above which the suspension of solute and dumbbell phase separates, which agrees with the numerical solution of the full eigenvalue problem. Our results show that the newly-identified couplings between the solute and the dumbbell distribution drive the phase separation. This mechanism is radically different from that driving the shear banding of micelles \cite{Fielding_2014,Germann_2019} and of entangled polymer solutions \cite{Cromer_2013,Cromer_2014,Peterson_2016,Peterson_2020,Burroughs_2021} and could act in parallel with the other two mechanisms. 

In summary, the model presented here is general and can be applied to study the flow of polymers that respond to different chemical stimuli such as local changes in pH, ionic strength or solute concentration changes. In future works, we plan to extend the framework to consider the finite extensibility of the spring, mean-field hydrodynamic interactions and conformation-dependent dumbbell friction coefficients. Finally, a suggestive application of the theory presented here concerns living systems whereby active processes, chemical reactions and inhomogeneous straining flows could be used to control the conformation and the spatial arrangements of macromolecules and proteins. 

\begin{acknowledgments}
Marco De Corato acknowledges funding from the European Union’s Horizon 2020 research and innovation program under the Marie Skłodowska-Curie action (GA 712754), the Severo Ochoa programme (SEV-2014-0425), the CERCA Programme/Generalitat de Catalunya and the MINECO through the Juan de la Cierva Incorporaci\'{o}n ICJ2018-035270-I. Marino Arroyo acknowledges the  support from the Generalitat de Catalunya (ICREA Academia Award, 2017-SGR-1278), from the Spanish Ministry of Economy and Competitiveness, through the Severo Ochoa Programme (CEX2018-000797-S), and from the European Research Council (CoG-681434). 
\end{acknowledgments}

\appendix
\section{Calculation of the rate of change of the free energy} \label{appFreeEn}
Here we report the details on the derivation of the rate of change of the free energy, which is obtained by taking the time derivative of Eq. \eqref{free_energ}:
\begin{multline}\label{free_energ_time_deriv1}
\mathcal{\dot{F}} = k_\text{B} T \int_\Omega (\log{c}+1)\frac{\partial c}{\partial t} \, d\Omega + k_\text{B} T \int_\Omega \int_{\boldsymbol{r}} (\log{\psi}+1)\frac{\partial \psi}{\partial t} \, d\Omega \, d\boldsymbol{r}+ \\ 
+\int_\Omega \int_{\boldsymbol{r}} \left(\frac{\partial \psi}{\partial t} U + \frac{\partial c}{\partial t} \frac{\partial U}{\partial c} \psi \right)  \, d\Omega \, d\boldsymbol{r}  \, \, ,
\end{multline}
where we used the fact that the integration domain does not change in time and we used the chain rule $\partial U/ \partial t = (\partial U/ \partial c) (\partial c/ \partial t)$.
We now substitute the time derivatives of $\psi$ and of $c$ from the continuity equations, given by Eqs. \eqref{cont_eq_dumbb}-\eqref{cont_eq_sol}, to obtain:
\begin{multline}\label{free_energ_time_deriv2}
\mathcal{\dot{F}} = -k_\text{B} T \int_\Omega (\log{c}+1) \frac{\partial}{\partial \boldsymbol{x}} \cdot (\boldsymbol{w} c ) \, d\Omega - k_\text{B} T \int_\Omega \int_{\boldsymbol{r}} (\log{\psi}+1)\left[ \frac{\partial}{\partial \boldsymbol{r}} \cdot (\dot{\boldsymbol{r}} \psi )+ \frac{\partial}{\partial \boldsymbol{x}} \cdot (\boldsymbol{w}_d \psi ) \right] \, d\Omega \, d\boldsymbol{r}+ \\
-\int_\Omega \int_{\boldsymbol{r}} \left[\left( \frac{\partial}{\partial \boldsymbol{r}} \cdot (\dot{\boldsymbol{r}} \psi )+ \frac{\partial}{\partial \boldsymbol{x}} \cdot (\boldsymbol{w}_\text{d} \psi ) \right) U + \left(\frac{\partial}{\partial \boldsymbol{x}} \cdot (\boldsymbol{w} c ) \right) \frac{\partial U}{\partial c} \psi \right]  \, d\Omega \, d\boldsymbol{r}  \, \, .
\end{multline}
We apply integration by parts to each of the integrals appearing in Eq. \eqref{free_energ_time_deriv2}. Since the velocity of dumbbells and of the solute normal to the boundaries vanishes and $\psi \rightarrow 0$ as $\boldsymbol{r} \rightarrow \infty$, the boundary terms coming from integration by parts disappear:
\begin{multline}\label{free_energ_time_deriv3}
\mathcal{\dot{F}} = k_\text{B} T \int_\Omega \frac{\partial}{\partial \boldsymbol{x}} \left(\log{c} \right) c \cdot \boldsymbol{w}  \, d\Omega +\int_\Omega \int_{\boldsymbol{r}} \frac{\partial}{\partial \boldsymbol{r}} ( k_\text{B} T \log{\psi}+U)\psi \cdot \dot{\boldsymbol{r}} \, d\Omega \, d\boldsymbol{r}+ \\
+ \int_\Omega \int_{\boldsymbol{r}} \frac{\partial}{\partial \boldsymbol{x}} (k_\text{B} T \log{\psi}+U) \cdot (\boldsymbol{w}_\text{d} \psi ) \, d\Omega \, d\boldsymbol{r}+\int_\Omega \int_{\boldsymbol{r}} \frac{\partial}{\partial \boldsymbol{x}} \left( \frac{\partial U}{\partial c} \psi  \right) \cdot \left(  \boldsymbol{w} c \right) \, d\Omega \, d\boldsymbol{r} \, \, .
\end{multline}
Finally, we use the property of the derivatives of the logarithms, which yields
\begin{multline}
\mathcal{\dot{F}} = k_\text{B} T \int_\Omega \frac{\partial c }{\partial \boldsymbol{x}} \cdot \boldsymbol{w} \, d\Omega  +\int_\Omega \int_{\boldsymbol{r}} \left( k_\text{B} T \frac{\partial \psi}{\partial \boldsymbol{r}} +\psi \frac{\partial}{\partial \boldsymbol{r}}U \right) \cdot \dot{\boldsymbol{r}} \, d\Omega \, d\boldsymbol{r}+ \\
+ \int_\Omega \int_{\boldsymbol{r}} \left(k_\text{B} T \frac{\partial \psi}{\partial \boldsymbol{x}}+\psi \frac{\partial}{\partial \boldsymbol{x}}U \right) \cdot \boldsymbol{w}_\text{d}  \, d\Omega \, d\boldsymbol{r}+\int_\Omega \int_{\boldsymbol{r}} \frac{\partial}{\partial \boldsymbol{x}} \left( \frac{\partial U}{\partial c} \psi  \right) \cdot \left(  \boldsymbol{w} c \right) \, d\Omega \, d\boldsymbol{r} \, \, .
\end{multline}

\section{Dissipation due to the dumbbells} \label{appA}
The dissipation is introduced by the relative motion between the two beads of the dumbbells and the local fluid velocity. Let's consider dumbbells whose center of mass lies at a position $\boldsymbol{x}$, and it is characterized by an end-to-end vector $\boldsymbol{r}$. The contribution to the total dissipation due to these dumbbells, $\mathcal{D}_d$, can be written as
\begin{multline}\label{dissipation_dumbb}
\mathcal{D}_d = \frac{\xi_\text{d}}{2} \int_{\Omega} \int_{\boldsymbol{r}} \, \psi \left[ \boldsymbol{w}_{d,1} - \boldsymbol{v}\left(\boldsymbol{x}+\frac{\boldsymbol{r}}{2}\right)\right]^2 \, d\Omega \, d\boldsymbol{r}+\frac{\xi_\text{d}}{2} \int_{\Omega} \int_{\boldsymbol{r}} \, \psi \left[ \boldsymbol{w}_{d,2} - \boldsymbol{v}\left(\boldsymbol{x}-\frac{\boldsymbol{r}}{2}\right)\right]^2 \, d\Omega \, d\boldsymbol{r}  \, \, ,
\end{multline}
where we wrote $\boldsymbol{w}_{d,1}$ as the velocity of the bead position at $\boldsymbol{x}+\boldsymbol{r}/2$ and $\boldsymbol{w}_{d,1}$ as the velocity of the bead position at $\boldsymbol{x}-\boldsymbol{r}/2$, $\xi_\text{d}$ is the drag coefficient of the beads. Assuming that the velocity field does not vary appreciably over the length of the dumbbell, we can Taylor expand the velocity field $\boldsymbol{v}(\boldsymbol{x}+\boldsymbol{r}/2)$ and $\boldsymbol{v}(\boldsymbol{x}-\boldsymbol{r}/2)$ around the position of the center of mass of the dumbbell $\boldsymbol{x}$
\begin{equation}
\boldsymbol{v}\left(\boldsymbol{x}\pm\frac{\boldsymbol{r}}{2}\right) \approx \boldsymbol{v}(\boldsymbol{x}) \pm \frac{1}{2} \boldsymbol{r} \cdot \frac{\partial}{\partial \boldsymbol{x}}\boldsymbol{v} \, \, ,
\end{equation}
where the gradient of the velocity field is evaluated at the center of mass of the dumbbell $\boldsymbol{x}$. Substituting this into the definition of the dissipation, given by Eq. \eqref{dissipation_dumbb}, and carrying out the square, we obtain:
\begin{multline}\label{dissipation_dumbb1}
\mathcal{D}_\text{d} = \frac{\xi_\text{d}}{2} \int_{\Omega} \int_{\boldsymbol{r}} \, \psi \bigg[ \boldsymbol{w}_{\text{d},1}^2 + \boldsymbol{w}_{d,1}^2 -2 \boldsymbol{w}_{\text{d},1} \cdot \boldsymbol{v} -2 \boldsymbol{w}_{\text{d},2} \cdot \boldsymbol{v} +  \boldsymbol{v}^2 + \\
+ \frac{1}{2} \left( \boldsymbol{r} \cdot \frac{\partial}{\partial \boldsymbol{x}}\boldsymbol{v}\right)^2 - \frac{1}{2} \boldsymbol{w}_{\text{d},1} \cdot \left( \boldsymbol{r} \cdot \frac{\partial}{\partial \boldsymbol{x}}\boldsymbol{v}\right) + \frac{1}{2} \boldsymbol{w}_{\text{d},2} \cdot \left( \boldsymbol{r} \cdot \frac{\partial}{\partial \boldsymbol{x}}\boldsymbol{v} \right) \bigg] \, d\Omega \, d\boldsymbol{r}  \, \, ,
\end{multline}
where the velocity and the gradient of the velocity are evaluated at the dumbbell center of mass $\boldsymbol{x}$.
Considering that the velocity of the center of mass, $\boldsymbol{w}_d$, can be written as
\begin{equation}
\boldsymbol{w}_\text{d} = \frac{1}{2} \left(\boldsymbol{w}_{\text{d},1}+\boldsymbol{w}_{\text{d},2} \right) \, \, ,
\end{equation}
and the rate of change of the end-to-end vector, $\dot{\boldsymbol{r}}$,  can be written as
\begin{equation}
\dot{\boldsymbol{r}} = \left(\boldsymbol{w}_{\text{d},1}-\boldsymbol{w}_{\text{d},2} \right) \, \, ,
\end{equation}
we rewrite Eq. \eqref{dissipation_dumbb1} in terms of $\boldsymbol{w}_d$ and $\dot{\boldsymbol{r}}$: 
\begin{equation}\label{dissipation_dumbb2}
\mathcal{D}_\text{d} = \xi_\text{d} \int_{\Omega} \int_{\boldsymbol{r}} \, \psi \left( \boldsymbol{w}_\text{d} - \boldsymbol{v}\boldsymbol{x}\right)^2 \, d\Omega \, d\boldsymbol{r}+\frac{\xi_\text{d}}{4} \int_{\Omega} \int_{\boldsymbol{r}} \, \psi \left[ \dot{\boldsymbol{r}} - \boldsymbol{r} \cdot \frac{\partial}{\partial \boldsymbol{x}}\boldsymbol{v} \right]^2 \, d\Omega \, d\boldsymbol{r}  \, \, ,
\end{equation}
which is the dumbbell contribution to the dissipation used in the Eq. \eqref{dissipation} of the main text.

\section{perturbation of the free energy around equilibrium}\label{app_pert_eq}
In this Appendix, we analyze the stability of a homogeneous solution of solute and dumbbells at equilibrium. We consider perturbations to the solute and dumbbell distributions and we check if these perturbations result in a reduction of the total free energy of the system. If this is the case, then small perturbations will lead to phase separation.
The free energy of the system is defined in the main text as
\begin{equation}
\mathcal{F} = k_\text{B} T \int_\Omega c \log{c} \, d\Omega + k_\text{B} T \int_\Omega \int_{\boldsymbol{r}} \psi \log{\psi} \, d\Omega \, d\boldsymbol{r}+\int_\Omega \int_{\boldsymbol{r}} \psi U  \, d\Omega \, d\boldsymbol{r}  \, \, ,
\end{equation}
Where the elastic energy is given by $U=\frac{1}{2} k(c) \vert\boldsymbol{r}\vert^2$.

We consider perturbations around equilibrium of the type:
\begin{equation}\label{perturb_freeng}
c = c_{eq}(1+\delta c) \;\;\,  \psi = \psi_{eq}(1+\delta \psi) \, \, ,
\end{equation}
where $c_{eq}$ is a constant and $\psi_{eq}$ is constant in space but it depends on the magnitude of the end to end distance vector:
\begin{equation}\label{psieq}
\psi_{eq}=\frac{n_\text{d,eq}}{2 \sqrt{2}}\left(\frac{k_\text{eq}}{\pi \, k_\text{B}T}\right)^{3/2} \, \exp{\left(-\frac{\vert\boldsymbol{r}\vert^2 }{2} \, \frac{k_\text{eq}}{k_\text{B}T}\right)} \, \, .
\end{equation}
As we showed in the main text, it is: 
\begin{equation}
 \int_{\boldsymbol{r}} \psi_{eq} \, d\boldsymbol{r} = n_\text{d,eq} \, \, .
 \end{equation}
 
By inserting the perturbations, given by Eqs. \eqref{perturb_freeng}, into the free energy and retaining only the second-order terms, we obtain the perturbation to the free energy:
\begin{multline}\label{free_energ1}
\delta \mathcal{F} = \frac{k_\text{B} T}{2} c_{eq} \int_\Omega \delta c^2 \, d\Omega + \frac{k_\text{B} T}{2}  \int_\Omega \int_{\boldsymbol{r}} \psi_{eq} \delta \psi^2 \, d\Omega \, d\boldsymbol{r}+\frac{c_{eq}^2}{4} \int_\Omega \int_{\boldsymbol{r}} \psi_{eq} \frac{d^2 k(c)}{d c^2}\Bigr|_{c_\text{eq}}   \delta c^2  \vert\boldsymbol{r}\vert^2 \, d\Omega \, d\boldsymbol{r} +\\
+\frac{c_{eq}}{2} \int_\Omega \int_{\boldsymbol{r}} \psi_{eq} \frac{d k(c)}{d c}\Bigr|_{c_\text{eq}}   \delta \psi \, \delta c \vert\boldsymbol{r}\vert^2 \, d\Omega \, d\boldsymbol{r}  \, \, .
\end{multline}
The system is unstable if a perturbation $\delta c$ or $\delta \psi$ leads to a negative free energy change $\delta \mathcal{F}<0$
To check this, we proceed by expand the perturbations fields in eigenfunctions. The expansion for $\delta c$ is given by
\begin{equation}\label{eigenexp1}
\delta c= \sum_{\boldsymbol{q}}  c_{\boldsymbol{q}} \, e^{ i \boldsymbol{q} \cdot \boldsymbol{x}} \, \, .
\end{equation}
In the equation above $\boldsymbol{q}$  is the wavevector defines as $\boldsymbol{q}=\frac{2\, \pi}{L} \{ n_x, n_y, n_z \}$ with $n_x$, $n_y$ and $n_z$ $\in \, \mathbb{Z}$ and $L$ is the size of the domain. The amplitude of the perturbation at a given wavevector is set by $c_{\boldsymbol{q}}$. Since the perturbations are real-valued functions, the coefficients $ c_{\boldsymbol{q}}$ must fulfill the condition $ c_{-\boldsymbol{q}}= c_{\boldsymbol{q}}^*$ where $ c_{\boldsymbol{q}}^*$ is the complex conjugate of $ c_{\boldsymbol{q}}$.

We insert the decomposition of $\delta c$, given by  Eq. \eqref{eigenexp1}, into the first integral on the right-hand side of Eq. \eqref{free_energ1}:
\begin{equation}
\frac{k_\text{B} T}{2} c_{eq} \int_\Omega \delta c^2 \, d\Omega = \frac{k_\text{B} T}{2} c_{eq} \, L^3 \, \delta_{\boldsymbol{q} \boldsymbol{q}'}  c_{\boldsymbol{q}}^* c_{\boldsymbol{q}'} \, \, ,
\end{equation}
where $L^3$ is the volume of the spatial domain and $\delta_{\boldsymbol{q} \boldsymbol{q}'} $ is the Kronecker delta symbol. 
Similarly, by inserting the decomposition of $\delta c$ given by Eq. \eqref{eigenexp1} into the third integral on the right-hand side of the free energy perturbation, Eq. \eqref{free_energ1}, we find:
\begin{equation}
\frac{c_{eq}^2}{4} \int_\Omega \int_{\boldsymbol{r}} \psi_{eq} \frac{d^2 k(c)}{d c^2}\Bigr|_{c_\text{eq}}   \delta c^2  \, d\Omega \, d\boldsymbol{r} = \frac{ 3 n_{eq} k_\text{B} T}{4} \, V \, \frac{c_{eq}^2}{k_{eq}} \frac{d^2 k(c)}{d c^2}\Bigr|_{c_\text{eq}}  \delta_{\boldsymbol{q} \boldsymbol{q}'}  c_{\boldsymbol{q}}^* c_{\boldsymbol{q}'} \, \, ,
\end{equation}
where we used the fact that 
\begin{equation}
 \int_{\boldsymbol{r}} \psi_{eq}   \vert\boldsymbol{r}\vert^2 d\boldsymbol{r} = \frac{ 3 n_\text{d,eq} k_\text{B} T}{k_{eq}} \, \, .
 \end{equation}

Now we need to assume an expansion for the perturbation of the dumbbell distribution, $\delta \psi$, which depends not only on space but also on the end-to-end space $\boldsymbol{r}$. We assume an expansion of the type:
\begin{equation}
\delta \psi= \sum_{\boldsymbol{q}} \sum_{j}  \Psi_{\boldsymbol{q},j} \,  e^{ i \boldsymbol{q} \cdot \boldsymbol{x}} \, \, \text{L}^{\frac{1}{2}}_j \left(\frac{r^2 k_BT}{2 k_{eq}} \right) \, \, ,
\end{equation}
where $\text{L}^{\frac{1}{2}}_j $ is the associated Laguerre polynomial of order $j$. The value $\Psi_{\boldsymbol{q},j}$ gives the amplitude of the perturbation with wavelength $\boldsymbol{q}$ and mode $j$. Basically, we are assuming that $\delta \psi$ can be separated in the product of two functions, one that only depends on space and one that only depends on the end-to-end vector $r$, we then choose the appropriate basis for these functions.  For the end-to-end vector space we choose the Laguerre polynomials because they have the following orthogonality property
\begin{equation}\label{eigexp2}
\int_0^\infty \text{L}^{\frac{1}{2}}_i(y) \text{L}^{\frac{1}{2}}_j(y)  e^{-y} y^{\frac{1}{2}}\, \, dy=\frac{ \Gamma(i+\frac{1}{2}+1)}{i!} \delta_{ij} \, \, .
\end{equation}
Since the perturbations are real valued, the coefficients $\Psi_{\boldsymbol{q},j}$ must fulfill the condition $\Psi_{-\boldsymbol{q},j}=\Psi_{\boldsymbol{q},j}^*$.
Finally, the expansion in Laguerre polynomials ensures that the perturbations, $\psi_{eq} \delta \psi$, decay to zero as $r\rightarrow \infty$ because $\psi_{eq} \propto \exp{(-r^2)}$. Therefore, they fulfill the boundary conditions at $r \rightarrow \infty$.

By inserting the expansion given by Eq. \eqref{eigexp2} into the second integral on the right-hand side of Eq. \eqref{free_energ1}, we obtain:
\begin{equation}
\frac{k_\text{B} T}{2}  \int_\Omega \int_{\boldsymbol{r}} \psi_{eq} \delta \psi^2 \, d\Omega \, d\boldsymbol{r}= 2 \, \pi k_\text{B} T \, \sum_{\boldsymbol{q}} \sum_{j}  \Psi_{\boldsymbol{q},j} \Psi_{\boldsymbol{q}',m}  \,   \int_\Omega e^{ i \boldsymbol{q} \cdot \boldsymbol{x}} \, e^{ i \boldsymbol{q}' \cdot \boldsymbol{x}} d\Omega \int_0^\infty  \psi_{eq} \text{L}^{\frac{1}{2}}_j \left(\frac{r^2 k_BT}{2 k_{eq}} \right) \text{L}^{\frac{1}{2}}_m \left(\frac{r^2 k_BT}{2 k_{eq}} \right) r^2 dr \, \, ,
\end{equation}
where we have rewritten the integral over $\boldsymbol{r}$ in spherical coordinates and we have carried out the integration over the theta and phi coordinates. We substitute the expression for $\psi_{eq} $ in the right-hand side to obtain 
\begin{multline}
\frac{k_\text{B} T}{2}  \int_\Omega \int_{\boldsymbol{r}} \psi_{eq} \delta \psi^2 \, d\Omega \, d\boldsymbol{r}= 2 \, \pi k_\text{B} T \, \frac{n_\text{d,eq}}{2 \sqrt{2}}\left(\frac{k_\text{eq}}{\pi \, k_\text{B}T}\right)^{3/2}  \,  \sum_{\boldsymbol{q}} \sum_{j}  \Psi_{\boldsymbol{q},j} \Psi_{\boldsymbol{q}',m}  \,   \int_\Omega e^{ i \boldsymbol{q} \cdot \boldsymbol{x}} \, e^{ i \boldsymbol{q}' \cdot \boldsymbol{x}} d\Omega \\
 \int_0^\infty  \, \exp{\left(-\frac{\vert\boldsymbol{r}\vert^2 }{2} \, \frac{k_\text{eq}}{k_\text{B}T}\right)} \text{L}^{\frac{1}{2}}_j \left(\frac{r^2 k_BT}{2 k_{eq}} \right) \text{L}^{\frac{1}{2}}_m \left(\frac{r^2 k_BT}{2 k_{eq}} \right) r^2 dr \, \, .
\end{multline}
By considering the change of variable $z=\frac{r^2 k_{eq}}{2 k_BT}$, the integral can be rewritten as 
\begin{multline}
\frac{k_\text{B} T}{2}  \int_\Omega \int_{\boldsymbol{r}} \psi_{eq} \delta \psi^2 \, d\Omega \, d\boldsymbol{r}= \\
k_\text{B} T \, \frac{n_\text{d,eq}}{\sqrt{\pi}} \,  \sum_{\boldsymbol{q}} \sum_{j}  \Psi_{\boldsymbol{q},j} \Psi_{\boldsymbol{q}',m}  \,   \int_\Omega e^{ i \boldsymbol{q} \cdot \boldsymbol{x}} \, e^{ i \boldsymbol{q}' \cdot \boldsymbol{x}} d\Omega \int_0^\infty  \, \exp{\left(-z\right)} \text{L}^{\frac{1}{2}}_j (z) \text{L}^{\frac{1}{2}}_m (z) z^{1/2} dz \, \,.
\end{multline}
These integrals can be carried out to give:
\begin{equation}
\frac{k_\text{B} T}{2}  \int_\Omega \int_{\boldsymbol{r}} \psi_{eq} \delta \psi^2 \, d\Omega \, d\boldsymbol{r}=k_\text{B} T \, L^3 \, \frac{n_\text{d,eq}}{\sqrt{\pi}} \,   \Psi_{\boldsymbol{q},j}^* \Psi_{\boldsymbol{q}',m}  \,  \frac{ \Gamma(j+\frac{1}{2}+1)}{j!} \delta_{jm} \delta_{\boldsymbol{q} \boldsymbol{q}'} 
\end{equation}
Now, the last integral on the right hand side can be evaluate using the same expansion for of the perturbations
\begin{multline}
\frac{c_{eq}}{2} \int_\Omega \int_{\boldsymbol{r}} \psi_{eq} \frac{d k(c)}{d c}\Bigr|_{c_\text{eq}}   \delta \psi \, \delta c \vert\boldsymbol{r}\vert^2 \, d\Omega \, d\boldsymbol{r} =\\
 2 \, \pi  \, c_{eq} \,  \frac{n_\text{d,eq}}{2 \sqrt{2}}\left(\frac{k_\text{eq}}{\pi \, k_\text{B}T}\right)^{3/2}  \, \frac{d k(c)}{d c}\Bigr|_{c_\text{eq}} \,   \sum_{\boldsymbol{q}} \sum_{j}  \Psi_{\boldsymbol{q},j}  c_{\boldsymbol{q}'}  \int_\Omega e^{ i \boldsymbol{q} \cdot \boldsymbol{x}} \, e^{ i \boldsymbol{q}' \cdot \boldsymbol{x}} d\Omega \int_0^\infty  \, \exp{\left(-\frac{\vert\boldsymbol{r}\vert^2 }{2} \, \frac{k_\text{eq}}{k_\text{B}T}\right)} \text{L}^{\frac{1}{2}}_j \left(\frac{r^2 k_BT}{2 k_{eq}} \right)  r^4 dr  \, \, .
\end{multline}
By using again the change of variable $z=\frac{r^2 k_{eq}}{2 k_BT}$, the integrals on the right-hand side can be rewritten as 
\begin{multline}
\frac{c_{eq}}{2} \int_\Omega \int_{\boldsymbol{r}} \psi_{eq} \frac{d k(c)}{d c}\Bigr|_{c_\text{eq}}   \delta \psi \, \delta c \vert\boldsymbol{r}\vert^2 \, d\Omega \, d\boldsymbol{r} = \\
2 \, \, c_{eq} \,  \frac{n_\text{d,eq}}{\sqrt{\pi}}\frac{k_\text{B}T}{k_\text{eq} }  \, \frac{d k(c)}{d c}\Bigr|_{c_\text{eq}} \,   \sum_{\boldsymbol{q}} \sum_{j}  \Psi_{\boldsymbol{q},j}  \, c_{\boldsymbol{q}'}  \int_\Omega e^{ i \boldsymbol{q} \cdot \boldsymbol{x}} \, e^{ i \boldsymbol{q}' \cdot \boldsymbol{x}} d\Omega \int_0^\infty  \, \exp{\left(-z\right)} \text{L}^{\frac{1}{2}}_j \left(z \right)  z^\frac{3}{2} dz  \, \, .
\end{multline}
The integrals on the right-hand side can be carried out to obtain:
\begin{multline}
\frac{c_{eq}}{2} \int_\Omega \int_{\boldsymbol{r}} \psi_{eq} \frac{d k(c)}{d c}\Bigr|_{c_\text{eq}}   \delta \psi \, \delta c \vert\boldsymbol{r}\vert^2 \, d\Omega \, d\boldsymbol{r} = 3 \,  L^3 \, c_{eq} \,  \frac{n_\text{d,eq}}{4}\frac{k_\text{B}T}{k_\text{eq} }  \, \frac{d k(c)}{d c}\Bigr|_{c_\text{eq}} \,   \left( \Psi_{\boldsymbol{q},j}^*  c_{\boldsymbol{q}'}+\Psi_{\boldsymbol{q}',j}  c_{\boldsymbol{q}}^* \right)  \delta_{\boldsymbol{q} \boldsymbol{q}'} \left(\delta_{j0} -\delta_{j1} \right) \, \, .
\end{multline}

The perturbations to the free energy can be expressed in matrix form as  
\begin{equation}
\delta \mathcal{F} = L^3 \, k_BT \, c_{eq} \, p^* A p
\end{equation} 
where the vectors $p$ and $p^*$ are given by
\begin{equation}
p=\left(  
\begin{matrix}
\delta c_{\boldsymbol{q}} \\
\Psi_{\boldsymbol{q},0}\\
\Psi_{\boldsymbol{q},1}\\
\vdots \\
\Psi_{\boldsymbol{q},j}
\end{matrix} \right) \, ; \, \, \, 
p^*=\left(  
\begin{matrix}
\delta c_{\boldsymbol{q}}^* \\
\Psi_{\boldsymbol{q},0}^*\\
\Psi_{\boldsymbol{q},1}^*\\
\vdots \\
\Psi_{\boldsymbol{q},j}^*
\end{matrix} \right)
\end{equation}
 and the matrix A is given by
\begin{equation}
A=
\left(
\begin{matrix}
\frac{1}{2} +  \frac{ 3 n_{eq} }{4 c_{eq}} \, \frac{c_{eq}^2}{k_{eq}} \frac{d^2 k(c)}{d c^2}\Bigr|_{c_\text{eq}}&  \frac{3 n_\text{d,eq}}{4 c_{eq}}\frac{c_{eq}}{k_\text{eq} }  \, \frac{d k(c)}{d c}\Bigr|_{c_\text{eq}}  & -\frac{3 n_\text{d,eq}}{4 c_{eq}}\frac{c_{eq}}{k_\text{eq} }  \, \frac{d k(c)}{d c}\Bigr|_{c_\text{eq}}  & 0 & 0 \\
\frac{3 n_\text{d,eq}}{4 c_{eq}}\frac{c_{eq}}{k_\text{eq} }  \, \frac{d k(c)}{d c}\Bigr|_{c_\text{eq}} & \frac{n_\text{d,eq}}{2 c_{eq}} &0 & 0 & 0 \\
-\frac{3 n_\text{d,eq}}{4 c_{eq}}\frac{c_{eq}}{k_\text{eq} }  \, \frac{d k(c)}{d c}\Bigr|_{c_\text{eq}} & 0 & \frac{3 n_\text{d,eq}}{4 c_{eq}} & 0 & 0\\
0 & 0 & 0& \ddots & 0 \\
0 & 0 & 0 & 0 & \, \frac{n_\text{d,eq}}{\sqrt{\pi}  \, c_{eq}} \,  \frac{ \Gamma(j+\frac{1}{2}+1)}{j!}
\end{matrix} \right)
\end{equation} 
The homogeneous equilibrium system is unstable if the determinant of the $A$ is negative $\text{det(}A)<0$. 
The condition of stability can be rewritten 
\begin{equation}
-\alpha^2  \left( \frac{15}{4}\alpha \, \beta^2 -\frac{3}{2} \alpha \, \gamma  -1\right)<0 \, \, ,
\end{equation}
where, following the nomenclature used in the main text, we introduced the dimensionless groups $\alpha=\frac{ n_{eq} }{c_{eq}}$, $\beta=\frac{c_{eq}}{k_\text{eq} }  \, \frac{d k(c)}{d c}\Bigr|_{c_\text{eq}} $ and $\gamma= \frac{c_{eq}^2}{k_{eq}} \frac{d^2 k(c)}{d c^2}\Bigr|_{c_\text{eq}}$.
The critical stability is given when the determinant of the matrix is exactly zero
\begin{equation}
-\alpha^2  \left( \frac{15}{4}\alpha \, \beta^2 -\frac{3}{2} \alpha \, \gamma  -1\right)=0 \, \, .
\end{equation}
The condition is satisfied when
\begin{equation}
\left( \frac{15}{4}\alpha \, \beta^2 -\frac{3}{2} \alpha \, \gamma  -1\right)=0 \, \, ,
\end{equation}
which leads to the relation 
\begin{equation}
\alpha =  \frac{4} {15 \, \beta^2 -6  \, \gamma  } \, \, .
\end{equation}
This relation describes a curve in the parameter space that discriminates regions that are stable from regions that are unstable.
In the main text we show that this expression matches the numerical results obtained from the linear stability analysis.

\section{Governing equations using the closure approximation and the Einstein notation} \label{AppEinst}
In this appendix we rewrite Eqs. \eqref{conftenstransp2} - \eqref{solbal2} using Einstein index notation where repeated index are summed. We assume that a cartesian coordinate system is employed.
The equation for the local conformation tensor, $C_{ij}$, reads:
\begin{equation}
\overset{\nabla}{C_{ij}}  = -\frac{4}{\xi_\text{d}}\left( k(c) C_{ij}-k_\text{B}T n_\text{d} \delta_{ij}\right) + \frac{k_\text{B} T}{2 \xi_\text{d}} \frac{\partial }{\partial x_k} \left(\frac{\partial }{\partial x_k} C_{ij} \right)+ \frac{5}{12 \xi_\text{d}} \frac{\partial }{\partial x_l} \left( \frac{d k(c)}{d c} \frac{\partial c}{\partial x_l} \frac{C_{mm}}{\Psi} C_{ij} \right) \, \, ,
\end{equation}
where 
\begin{equation}
\overset{\nabla}{C_{ij}} = \frac{\partial C_{ij}}{\partial t} + v_k  \frac{\partial }{\partial x_k} C_{ij}- \left( \frac{\partial }{\partial x_l }  v_i \right) C_{lj}-C_{im}  \frac{\partial }{\partial x_m}v_j\, \, ,
\end{equation}
denotes the upper convected derivative of the tensor $C_{ij}$ and $ \delta_{ij}$ represents the Kronecker delta.

The equation for the local dumbbell density, $n_\text{d}$, rewritten using the Einstein notation reads:
\begin{equation}
\frac{\partial n_\text{d}}{\partial t} + v_i  \frac{\partial n_\text{d}}{\partial x_i} = \frac{k_\text{B} T}{2 \xi_\text{d}} \frac{\partial }{\partial x_j} \left(\frac{\partial }{\partial x_j} n_\text{d}\right)  + \frac{1}{4 \xi_\text{d}} \frac{\partial }{\partial x_k} \cdot \left( \frac{d k(c)}{d c} \frac{\partial c}{\partial x_k} C_{ll}\right) \, \, .\end{equation}
The momentum balance reads:
\begin{equation}
 \frac{\partial }{\partial x_k} \left[ 2 \eta \, D_{ki} - P \delta_{ki} +   k(c)C_{ki}-k_\text{B} T n_\text{d} \delta_{ki}   \right ]=0 \,  \, ,
\end{equation} 
The equation that governs the evolution of the solute using the Einstein notation:
\begin{equation}
\frac{\partial c}{\partial t}  = \frac{\partial }{\partial x_k} \cdot \left( \frac{k_\text{B} T}{\xi} \frac{\partial c}{\partial x_k} +\frac{c}{2 \xi} \frac{\partial}{\partial x_k} \left( \frac{d k(c)}{d c} \, C_{ii} \right) - c v_k\right)\, \, ,
\end{equation} 

\section{Details on the linear stability analysis in the case of homogeneous shear}\label{appB}
The linear stability analysis proceeds by substituting the Fourier expansions given by Eqs \eqref{fourierconc}-\eqref{fouriervel} into Eqs \eqref{perturbshearconcsol}-\eqref{perturbshearxmombal}. By substituting the expansion into the momentum balance we obtain:
\begin{multline}
- \eta^* \sum_{n=1}^{\infty} (2 \, \pi \, n)^2 v^{*'}_{x,n} \sin{(2 \pi n y^*)} \, \exp{(\lambda_n \, t)} - \frac{1}{4 \, De} \sum_{n=1}^{\infty}  (2 \, \pi \, n)  C^{*'}_{xy,n} \sin{(2 \pi n y^*)} \, \exp{(\lambda_n \, t)} +\\
-\frac{1}{4} \frac{\partial k^*(c^*)}{\partial c^*}\Bigr|_{c^*=1} \, \sum_{n=1}^{\infty}  (2 \, \pi \, n)  c^{*'}_{n} \sin{(2 \pi n y^*)} \, \exp{(\lambda_n \, t)} = 0 \, \, ,
\end{multline}
which can be turned into an equation for the coefficients
\begin{equation}\label{velpertfourier}
- \eta^* (2 \, \pi \, n) v^{*'}_{x,n} - \frac{1}{4 \, De}  C^{*'}_{xy,n} -\frac{1}{4} \frac{d k^*(c^*)}{d c^*}\Bigr|_{c^*=1} \,  c^{*'}_{n} = 0 \, \, .
\end{equation}
We proceed similarly for the equation of transport of solute, dumbbells and conformation tensor. Substituting the Fourier expansion into the equation of transport of solute and equating term by term yields
\begin{multline}\label{solpertfourier}
\left[\lambda_n +\frac{2 \, \pi^2 \, n^2}{Pe} \left(2 + n_0^* (3 + 2\, De^2) \frac{d^2 k^*(c^*)}{d c^{*2}}\Bigr|_{c^*=1} \right) \right] c^{*'}_n = \\
=- n_0^* \frac{2 \, \pi^2 \, n^2}{Pe} \frac{d k^*(c^*)}{d c^*}\Bigr|_{c^*=1} \left(C^{*'}_{xx,n}+C^{*'}_{yy,n}+C^{*'}_{zz,n} \right) \, \, .
\end{multline}
The Fourier modes of the perturbation to the dumbbell number density satisfy
\begin{equation}
\left(\lambda_n +\frac{2 \, \pi^2 \, n^2}{Pe_d} \right) n^{*'}_{\text{d},n} = - n_0^* \frac{2 \, \pi^2 \, n^2}{Pe_d} (3 + 2\, De^2) \frac{d k^*(c^*)}{d c^*}\Bigr|_{c^*=1} c^{*'}_n \, \, .
\end{equation}
The Fourier modes for each component of the conformation tensor are given by
\begin{multline}
\left(\lambda_n +\frac{2 \, \pi^2 \, n^2}{Pe_p} +\frac{1}{De} \right) C^{*'}_{xx,n} + \frac{1}{2 \, \eta^*}\left(C^{*'}_{xy,n}+\frac{d k^*(c^*)}{d c^*}\Bigr|_{c^*=1} c^{*'}_n \right) - 2 C^{*'}_{xy,n} = \\
=  \frac{1}{De} \, n^{*'}_{\text{d},n} - \frac{1}{De} \frac{d k^*(c^*)}{d c^*}\Bigr|_{c^*=1} (1+2 De^2) c^{*'}_n -\frac{5 \, \pi^2 \, n^2}{3 \, Pe_p} (1+2 De^2)(3+2De^2) \frac{d k^*(c^*)}{d c^*}\Bigr|_{c^*=1} c^{*'}_n \, \, ,
\end{multline}
where the second, third and fourth term in the left hand side come from the upper convected derivative and we have substituted the perturbations to the velocity as given by Eq. \eqref{velpertfourier}.
The Fourier modes of the yy and zz components are given by 
\begin{equation}
\left(\lambda_n +\frac{2 \, \pi^2 \, n^2}{Pe_p} +\frac{1}{De} \right) C^{*'}_{yy,n} =  \frac{1}{De} \, n^{*'}_{\text{d},n} - \frac{1}{De} \frac{d k^*(c^*)}{d c^*}\Bigr|_{c^*=1}  c^{*'}_n -\frac{5 \, \pi^2 \, n^2}{3 \, Pe_p} (3+2 De^2) \frac{d k^*(c^*)}{d c^*}\Bigr|_{c^*=1} c^{*'}_n \, \, ,
\end{equation}
\begin{equation}
\left(\lambda_n +\frac{2 \, \pi^2 \, n^2}{Pe_p} +\frac{1}{De} \right) C^{*'}_{zz,n} =  \frac{1}{De} \, n^{*'}_{\text{d},n} - \frac{1}{De} \frac{d k^*(c^*)}{d c^*}\Bigr|_{c^*=1}  c^{*'}_n -\frac{5 \, \pi^2 \, n^2}{3 \, Pe_p} (3+2 De^2) \frac{d k^*(c^*)}{d c^*}\Bigr|_{c^*=1} c^{*'}_n \, \, .
\end{equation}
The Fourier modes of the xy component are given by
\begin{multline}\label{shearconfpertfourier}
\left(\lambda_n +\frac{2 \, \pi^2 \, n^2}{Pe_p} +\frac{1}{De}+ \frac{1}{4 \, De \eta^*} \right) C^{*'}_{xy,n} - C^{*'}_{yy,n} = \\
= \left(- \frac{d k^*(c^*)}{d c^*}\Bigr|_{c^*=1}  -\frac{5 \, \pi^2 \, n^2}{3 \, Pe_p} De \, (3+2De^2) \frac{d k^*(c^*)}{d c^*}\Bigr|_{c^*=1} -\frac{1}{4 \, De \eta^*} \right) c^{*'}_n\, \, .
\end{multline}
Eqs. \eqref{solpertfourier}-\eqref{shearconfpertfourier} represent a linear system, which can be rewritten as
\begin{equation}
\lambda_n \, X - E \cdot X =0 \, \, ,
\end{equation}
where $X$ is a vector containing the inital values of the perturbations 
\begin{equation} \label{eigenproblemmatrix}
X = \begin{bmatrix} c^{*'}_n\\ n^{*'}_{\text{d},n} \\ C_{xx,n}^{*'} \\ C_{yy,n}^{*'} \\ C_{zz,n}^{*'} \\ C_{xy,n}^{*'} \end{bmatrix} \, \, ,
\end{equation}
and the matrix $E$ contains the coefficients that couple the same Fourier modes of different fields. The matrix $E$ is given by
\begin{equation}
\scalemath{0.7}{\begin{bmatrix} 
\frac{-2 (n \pi )^2}{Pe} \left(2 +n_0^* \, (3+ 2 \, De^2) \frac{d^2 k^*(c^*)}{d c^{*2}}\Bigr|_{c^*=1}   \right) & 0 & \frac{-2 (n \pi )^2}{Pe} \frac{d k^*(c^*)}{d c^*}\Bigr|_{c^*=1} & \frac{-2 (n \pi )^2}{Pe} \frac{d k^*(c^*)}{d c^*}\Bigr|_{c^*=1}  & \frac{-2 (n \pi )^2}{Pe} \frac{d k^*(c^*)}{d c^*}\Bigr|_{c^*=1} & 0 \\
  -n^* \frac{2 \, \pi^2 \, n^2}{Pe_d} (3 + 2\, De^2) \frac{d k^*(c^*)}{d c^*}\Bigr|_{c^*=1} & -\frac{2 \, \pi^2 \, n^2}{Pe_d} & 0 & 0  &0 & 0 \\
 -\frac{d k^*(c^*)}{d c^*}\Bigr|_{c^*=1} \left( n_0^* \frac{5 \, \pi^2 \, n^2}{3 \, Pe_d} (1 + 2\, De^2)(3 + 2\, De^2) +\frac{1+2\, De^2}{De} +\frac{1}{4 \, \eta^*} \right) & \frac{1}{De} &-\frac{1}{De} -\frac{-2 (n \pi )^2}{Pe_d}&0 &0 & 2- \frac{1}{2\, \eta^*} \\
 -\frac{d k^*(c^*)}{d c^*}\Bigr|_{c^*=1} \left( n_0^* \frac{5 \, \pi^2 \, n^2}{3 \, Pe_d} (3 + 2\, De^2) +\frac{1}{De} \right) & \frac{1}{De} &0 & -\frac{1}{De} -\frac{-2 (n \pi )^2}{Pe_d}& 0&  0\\
 -\frac{d k^*(c^*)}{d c^*}\Bigr|_{c^*=1} \left( n_0^* \frac{5 \, \pi^2 \, n^2}{3 \, Pe_d} (3 + 2\, De^2) +\frac{1}{De} \right)  & \frac{1}{De} & 0 & 0 & -\frac{1}{De}-\frac{-2 (n \pi )^2}{Pe_d}& 0\\
 -\frac{d k^*(c^*)}{d c^*}\Bigr|_{c^*=1} \left( n_0^* \frac{5 \, \pi^2 \, n^2}{3 \, Pe_d} De \, (3 + 2\, De^2) +\frac{1}{4 \, \eta^*} +1 \right) & 0 & 0 & 1 & 0 & -\frac{1}{De}-\frac{-2 (n \pi )^2}{Pe_d}-\frac{1}{4 \, De \, \eta^*}
\end{bmatrix} }
\end{equation}
The growth rate $\lambda_n$ that governs the stability of the homogeneous state to the perturbations is obtained by solving the linear system given by Eq. \eqref{eigenproblemmatrix}, which reduces to find the eigenvalues of the matrix $E$ for each Fourier mode $n$. 

\section{Derivation of the simplified expression for the critical Deborah number} \label{appC}
As discussed in the main text, in typical conditions, it is $Pe_d \gg 1$, $Pe/Pe_d \ll 1$, $De/Pe_d \ll 1$, $De/Pe \ll 1$ and $\eta^* \ll 1$. Therefore, we introduce the following simplifications to the linear stability analysis: (i) we neglect the perturbations of dumbbell density $n_\text{d}^{*'}=0$, since they evolve over a very slow timescale; (ii) we assume that the conformation tensor relaxes faster than any other timescale in the system; (iii) we take the limit $\eta^* \rightarrow 0$. Given these assumptions, the components of the conformation tensor are given by:
\begin{equation}
C^{*'}_{yy,n} = C^{*'}_{zz,n} = - \frac{d k^*(c^*)}{d c^*}\Bigr|_{c^*=1} c^{*'}_n \, \, ,
\end{equation}
\begin{equation}
C^{*'}_{xy,n} = - \frac{d k^*(c^*)}{d c^*}\Bigr|_{c^*=1} De \, c^{*'}_n \, \, ,
\end{equation}
\begin{equation}
C^{*'}_{xx,n} = - \frac{d k^*(c^*)}{d c^*}\Bigr|_{c^*=1} (1+ 2 \, De^2) c^{*'}_n \, \, .
\end{equation}
We substitute this expression in the equation for the evolution of the solute perturbations to obtain 
\begin{multline}
\left[\lambda_n +\frac{2 \, \pi^2 \, n^2}{Pe} \left(2 +16 \Lambda^* (n \, \pi)^2 + n_0^* (3 + 2\, De^2) \frac{d^2 k^*(c^*)}{d c^{*2}}\Bigr|_{c^*=1}\right) \right] c^{*'}_n = \\
= n_0^* \frac{2 \, \pi^2 \, n^2}{Pe} \left( \frac{d k^*(c^*)}{d c^*}\Bigr|_{c^*=1}\right)^2 \left(3 + 2\, De^2 \right) c^{*'}_n \, \, .
\end{multline}
We simplify $c^{*'}_n$ on the two sides of the equation to obtain a dispersion relation between the growth rate $\lambda_n$ and the dimensionless parameters of the problem.
\begin{equation}
\left[\lambda_n +\frac{2 \, \pi^2 \, n^2}{Pe} \left(2 + n_0^* (3 + 2\, De^2) \frac{d^2 k^*(c^*)}{d c^{*2}}\Bigr|_{c^*=1} \right) \right]  = n_0^* \frac{2 \, \pi^2 \, n^2}{Pe} \left( \frac{d k^*(c^*)}{d c^*}\Bigr|_{c^*=1}\right)^2 \left(3 + 2\, De^2 \right)  \, \, .
\end{equation}
We define a critical Deborah number, $De_\text{crit}$, as the value that leads to a marginally stable case, $\lambda_n =0$
\begin{equation}
 \left(2 + n_0^* (3 + 2\, De^2_\text{crit}) \frac{d^2 k^*(c^*)}{d c^{*2}}\Bigr|_{c^*=1} \right)  = n_0^*  \left( \frac{d k^*(c^*)}{d c^*}\Bigr|_{c^*=1}\right)^2 \left(3 + 2\, De^2_\text{crit} \right)  \, \, .
\end{equation}
Any homogeneous state with $De > De_\text{crit}$ leads to an instability and a shear-induced phase separation.
We can rearrange the equation above to write an explicit expression for $De_\text{crit}$
\begin{equation}
De_\text{crit}= \left( \frac{1 }{n_0^* \left[ \left( \frac{d k^*(c^*)}{d c^*}\Bigr|_{c^*=1} \right)^2 - \frac{d^2 k^*(c^*)}{d c^{*2}}\Bigr|_{c^*=1} \right]} - \frac{3}{2}\right)^{1/2} \, \, .
\end{equation}
%The right-hand side is minimum for $n=1$, meaning that the first Fourier mode to become unstable for $De>De_\text{crit}$ is the first one:
%\begin{equation}
%De_\text{crit}= \left( \frac{1}{n_0^* \left[ \left( \frac{d k^*(c^*)}{d c^*}\Bigr|_{c^*=1} \right)^2 - \frac{d^2 k^*(c^*)}{d c^{*2}}\Bigr|_{c^*=1} \right]} - \frac{3}{2}\right)^{1/2} \, \, .
%\end{equation}

%\nocite{*}

\bibliography{dumbbell_bib}% Produces the bibliography via BibTeX.

\end{document}